\documentclass[
 reprint,
 amsmath,amssymb,
 superscriptaddress,
 aps,
 prx,
 nofootinbib,
]{revtex4-2}

\usepackage{graphicx}
\usepackage{dcolumn}
\usepackage{bm}
\usepackage{enumitem,xcolor}
\usepackage{physics2}
\usephysicsmodule{ab}
\usephysicsmodule{op.legacy}
\usephysicsmodule{braket}
\usephysicsmodule{nabla.legacy}
\usepackage{derivative,fixdif}
\usepackage{mathtools}
\usepackage{multirow}
\usepackage{siunitx}
\usepackage{appendix}
\usepackage[colorlinks,linkcolor=blue,urlcolor=blue, citecolor=blue]{hyperref}
\usepackage[whole]{bxcjkjatype}
\usepackage[T1]{fontenc} 
\usepackage{comment}
\usepackage{amsfonts}

\setlist[itemize]{leftmargin=*}
\setlist[enumerate]{leftmargin=*}
\setlist[enumerate]{itemsep=-0mm}

\begin{document}

\title{Transversal surface-code game with reconfigurable qubits}
\author{Shinichi Sunami}
\email{shinichi.sunami@nano-qt.com}
\affiliation{Nanofiber Quantum Technologies, Inc. (NanoQT), 1-22-3 Nishiwaseda, Shinjuku-ku, Tokyo 169-0051, Japan.}
\affiliation{Clarendon Laboratory, University of Oxford, Oxford OX1 3PU, United Kingdom}
\author{Akihisa Goban}
\email{akihisa.goban@nano-qt.com}
\affiliation{Nanofiber Quantum Technologies, Inc. (NanoQT), 1-22-3 Nishiwaseda, Shinjuku-ku, Tokyo 169-0051, Japan.}
\author{Hayata Yamasaki}
\email{hayata.yamasaki@nano-qt.com}
\affiliation{Nanofiber Quantum Technologies, Inc. (NanoQT), 1-22-3 Nishiwaseda, Shinjuku-ku, Tokyo 169-0051, Japan.}
\affiliation{Department of Computer Science, Graduate School of Information Science and Technology, The University of Tokyo, 7-3-1 Hongo, Bunkyo-ku, Tokyo 113-8656, Japan}

\begin{abstract}

Qubit platforms supporting flexible connectivity have opened the door to novel theoretical advances in protocols for fault-tolerant quantum computation (FTQC), offering a compelling alternative to lattice surgery by leveraging transversal gates.
A wide variety of systems, such as atomic, photonic, and some solid-state qubits support such operations, among which neutral atoms have demonstrated notable progress enabled by large-scale parallel atom transport. 
However, a crucial gap remains between the theory of FTQC and its practical realization with these systems.
Most critically, a key theoretical requirement, that syndrome extraction must be performed frequently enough to keep error accumulation below a threshold constant, is not satisfied in a scalable manner in the existing approach, where increased transport times prevent meeting the required threshold conditions.
In this work, we develop a comprehensive theoretical framework that closes such a gap, bridging theoretical advances in fault-tolerant protocols with the latest experimental capabilities.
Building on the ``game of surface code'' framework originally developed for superconducting qubits, we introduce an alternative game-based paradigm for transversal-gate FTQC that harnesses the unique strengths of flexible qubit systems.
Focusing on surface code implementation, we formulate the FTQC protocol as a transversal surface-code game, governed by rules for mode transitions of code cells.
The rulesets are designed to enable syndrome extraction at any intermediate step during logical gate implementation, ensuring compatibility with the threshold theorem; at the same time, the framework fully exploits the capabilities of the hardware to scale and accelerate FTQC\@.
As a concrete application, our framework offers a systematic methodology and high-level abstraction for resource estimation and optimization in FTQC with improved space-time performance; an exemplary estimation for a neutral-atom system demonstrates performance competitive with a baseline lattice-surgery-based approach on superconducting qubits, even with physical operations on neutral atoms being orders of magnitude slower.
These results establish a solid foundation that bridges the theory and experiment of FTQC powered by reconfigurable qubit platforms, charting a well-founded pathway toward scalable, fault-tolerant quantum computers and setting practical directions for future technological development.
\end{abstract}
\maketitle

\section{Introduction}

The theory of fault-tolerant quantum computation (FTQC) lays the foundation for realizing large-scale quantum computers, promising broad applications in machine learning~\cite{lloyd2014quantum,  liu2021rigorous,NEURIPS2020_9ddb9dd5,yamasaki2022exponentialerrorconvergencedata,pmlr-v202-yamasaki23a,yamasaki2023advantagequantummachinelearning}, quantum chemistry~\cite{lee2021even, kim2022fault-tolerant}, cryptanalysis~\cite{shor1997polynomial-time, gidney2021how}, along with their impact on energy consumption~\cite{meier2025energy-consumption}.
A practical realization of FTQC must always be tailored to the characteristics of the underlying physical platform; as quantum technology advances, theoretical frameworks must also evolve to fully leverage new technological capabilities.
Recently, qubit platforms with flexible reconfigurability such as trapped ions and neutral atoms~\cite{Ryan-Anderson2024, henriet2020quantum, bluvstein2022quantum} have emerged as a leading candidate for FTQC, generating considerable excitement across quantum science and technology.
In particular, neutral atom arrays offer unique advantages, such as large-scale parallel reconfiguration to efficiently implement transversal gates; features that open new pathways to design efficient protocols for FTQC\@.
Among the many protocols for FTQC, surface code stands as the state-of-the-art, offering a high threshold for physical error rates and well-established procedures for logical gate implementation~\cite{fowler2012surface}.
Nevertheless, a substantial gap remains between the theory of FTQC and their practical realization in highly flexible qubit platforms; while the surface code and its variants have been extensively optimized for superconducting qubit platforms, a corresponding theoretical framework for neutral atom systems has yet to be fully established.

\begin{figure*}[t!]
    \centering
    \includegraphics[width=7.0in]{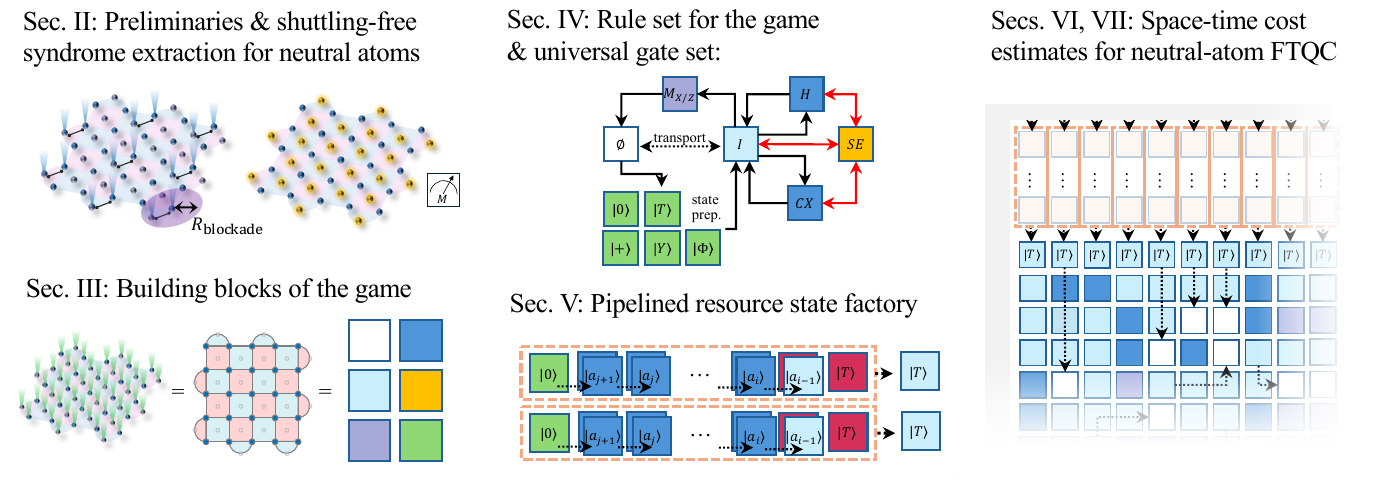}
    \caption{Outline of the paper.
    }
    \label{fig:outline}
\end{figure*}

Neutral atom platforms have demonstrated substantial experimental progress, including the realization of large-scale atom arrays~\cite{manetsch2024tweezer, norcia2024iterative, pichard2024rearrangement}, high-fidelity and parallel gate operations~\cite{bluvstein2024logical, bedalov2024fault-tolerant, reichardt2024logical}, and the integration of fast photonic interconnects for modular scaling~\cite{li2024high-rate, sunami2025scalable}.
Notably, the flexible reconfigurability of atom positions opens the door to new types of fault-tolerant protocols via transversal gates~\cite{Kubica2015,moussa2016transversal,chen2024transversal,cain2024correlated, zhou2024algorithmic, cain2025fast, sahay2025error, Turner2025, serraperalta2025}.
Provided that the physical error rate is below a certain threshold constant, fault-tolerant protocols using the surface code guarantee that the logical error rate can be arbitrarily suppressed, a celebrated result in the theory of FTQC known as the threshold theorem~\cite{10.1063/1.1499754,PhysRevLett.109.180502,10.5555/2685179.2685184,tamiya2024polylogtimeconstantspaceoverheadfaulttolerantquantum}.
Conventionally, such protocols assume an increasing number of measurement rounds for syndrome extraction at large scales.
In contrast, more recent theoretical advances in correlated decoding and algorithmic fault tolerance have established such threshold theorems using only a single measurement round of syndrome extraction with polynomial-time decoding, particularly for transversal logical gate implementations feasible with neutral atoms~\cite{zhou2024algorithmic, cain2024correlated, cain2025fast}.
These developments, together with the rapid development of decoding strategies for transversal logic~\cite{Turner2025, serraperalta2025}, have the potential to substantially accelerate FTQC~\cite{zhou2025resource}.
However, the theoretical guarantees of FTQC are valid only if the physical platform satisfies key requirements: most notably, the ability to perform syndrome extraction at least once within a constant time, ensuring that the physical error rate from decoherence and gate errors remains bounded below the threshold.

Fulfilling these requirements in the theory of FTQC presents fundamental technological challenges: reconfiguration over increasing distances leads to additional decoherence, shuttling-induced errors, and spatially varying gate times~\cite{bluvstein2024logical, manetsch2024tweezer, lam2021demonstration}.
Existing approaches, such as lattice surgery~\cite{horsman2012surface,litinski2019game}, may provide feasible solutions by using qubits fixed on a square-lattice layout, but sacrifice the unique advantages of movable qubits, failing to fully exploit reconfigurability or transversal gates and ultimately resulting in architectures that are merely slower versions of those based on superconducting qubits.
More recent proposals for long-range selective gates~\cite{pecorari2025high-rate, poole2025architecture, saffman2025quantum} partially address these limitations, but they are again limited by the increasing number of steps required for applying a sequence of long-range interactions in larger systems, losing the parallelism with increasing code distance~\cite{poole2025architecture,saffman2025quantum}.
As systems scale up, these effects become increasingly pronounced and threaten to violate the essential requirements of threshold theorems. This underscores the need to rethink the scalable design to ensure compatibility with the theory of FTQC\@.

Critically, a theoretical framework has been lacking for abstracting and optimizing transversal-gate FTQC protocols that leverage the advantages of flexible qubit platforms while remaining compatible with the fundamental requirements for FTQC\@.
For protocols based on lattice surgery, the ``game of surface code'' framework~\cite{litinski2019game} has established a crucial abstraction that bridges theoretical protocol design and physical implementation on superconducting qubits.
However, no analogous, fully general framework exists for transversal-gate, qubit-shuttling-enabled platforms.
Bridging this gap is essential not only to inform and guide experiments,
but also to advance the theoretical understanding of how the unique physical features can be harnessed to realize efficient, scalable FTQC that fulfills the requirements of threshold theorems.

In this work, we construct a comprehensive theoretical framework for the scalable implementation of the surface-code protocol to achieve universal quantum computation on platforms supporting transversal gates and mobile qubits, such as neutral atom arrays (see Fig.~\ref{fig:outline}).
Our approach identifies and clarifies optimized methods for FTQC in these systems, developing a suitable native instruction set to exploit the advantage of neutral atoms.
Inspired by the ``game of surface code'' abstraction for lattice surgery, we formulate the protocol for universal quantum computation with transversal gates as a game played on a two-dimensional grid, abstracting away physical details while explicitly connecting the experimental platform with the fault-tolerant protocol.
Within this approach, we introduce the concept of the cell: a rigid, mobile unit of the surface code with internal operational modes and mode-transition rules.
We describe how a finite rule set governs their transitions and transversal interactions on the grid to realize universal quantum computation.
A key innovation of our framework is the guarantee that syndrome extraction can be performed at any required timing of logical gate implementations, without waiting for time-consuming shuttling of atoms over large distances.
This enables the physical implementation to maintain the requirements of threshold theorems, even under realistic constraints of finite shuttling speed.
In particular, we propose that syndrome extraction should be performed using local selective gates and measurements without atom shuttling~\cite{radnaev2025universal,bedalov2024fault-tolerant}, a feature shared also by other qubit platforms with flexible reconfigurability~\cite{Moses2023,Sotirova2024low}.
Within this framework, we demonstrate that all logical operations can be efficiently implemented, while also providing the flexibility to extract syndromes at any intermediate time as necessary.
This ensures compatibility with the threshold theorems and, crucially, avoids the detours to designated zones that are required in conventional zoned approaches of neutral-atom implementations.
Enhanced efficiency stems from the use of transversal gates with a single round of shuttling-free syndrome extraction for algorithmic fault tolerance.

Further improvements are achieved by also optimizing gate teleportation and the preparation of required auxiliary states in the resource state factories, such as those for magic state distillation.
Previous works have made significant advances in the practical implementation of magic state distillation and in-place state preparation for superconducting qubits~\cite{litinski2019game, Litinski2019magicstate, gidney2024magic}; however, it has remained challenging to extend these techniques in a way that fully exploits the potential of transversal gates available on neutral atom platforms.
In contrast, we develop an alternative approach for designing space-time efficient factories by generalizing the methods in Ref.~\cite{litinski2019game}, originally developed for the lattice-surgery fault-tolerant protocol, to the setting of transversal-gate architectures.
Using Calderbank-Shor-Steane (CSS) codes, these factories can distill a variety of resource states required for gate teleportation, including those for implementing $T$ gates, $S$ gates, and remote CNOT gates across spatially separated modules.
Unlike the lattice-surgery-based factories, the transversal-gate-based factories proposed here are fully compatible with other key techniques in the transversal-gate architectures, such as algorithmic fault tolerance, and thus can be used with them to enhance space-time efficiency.
We also achieve additional optimization of such resource state factories through the pipelined implementation of distillation protocols and direct shuttling of qubits.

Moreover, we evaluate the performance of our approach through concrete resource estimates and time-cost analysis, including the example of implementing a $100$-qubit, $10^8$ $T$-count circuit using our framework.
Our results demonstrate for neutral-atom qubits that, with existing or foreseeable experimental parameters, this approach enables such quantum computation using on the order of $10^4$ physical qubits in several hours, a competitive estimation even considering other modalities.
The key to this capability is the exploitation of the unique physical features of neutral atom platforms, enabling surface-code FTQC to be realized substantially faster than lattice surgery by supporting the efficient implementation techniques described above.
Our framework makes resource estimation more systematic and scalable than the naive method that directly rewrites a fault-tolerant circuit into a sequence of native physical operations, since we provide an abstraction in the form of a game that bridges theory and experiment.
This abstraction is also expected to facilitate further quantitative analysis and optimization in future technological developments.

Consequently, these results advance both the theory and practical implementation of FTQC, particularly based on the progress of reconfigurable qubit technologies.
Our framework provides a theoretical foundation that establishes a previously missing bridge between the requirements of the threshold theorem and their experimental realization with transversal gates, identifies scalable design principles for the neutral atom platform, and sets technological goals for future experimental development.
By clarifying and abstracting the rules of the ``transversal surface-code game,'' our approach opens new avenues for FTQC, enabling fault-tolerant protocols and implementations that fully leverage the unique physical features, capabilities previously inaccessible with conventional theoretical 
approaches such as lattice surgery.

This paper is organized as follows.
In Sec.~\ref{sec:prelim}, we first provide a brief introduction to neutral-atom qubits and surface-code fault-tolerant quantum computation, and then discuss shuttling-free syndrome extraction that will be the basis of our architecture.
We then describe the components and rules of the game: in Sec.~\ref{sec:logical-cell}, we first define the cell, and in Sec.~\ref{sec:ruleset}, we describe a ruleset to ensure fault tolerance in the logical-cell operations without referring to the specifics of neutral-atom implementation.
We discuss a pipelined resource state factory design for our architecture in Sec.~\ref{sec:factory} and illustrate the factories for magic state distillation.
In Sec.~\ref{sec:operating}, we describe a universal instruction set for a neutral-atom array compatible with our requirements, and discuss its predicted performance.
Based on the quantitative analysis, in Sec.~\ref{sec:example}, we perform a baseline evaluation of the speed of the neutral-atom FTQC executing generic Clifford+$T$ circuits, and in Sec.~\ref{sec:concl} we conclude the paper and provide an outlook to scale the proposed architecture further.

\section{Surface-code FTQC with neutral atoms}\label{sec:prelim}

\begin{figure*}[t]
    \centering
    \includegraphics[width=7.0in]{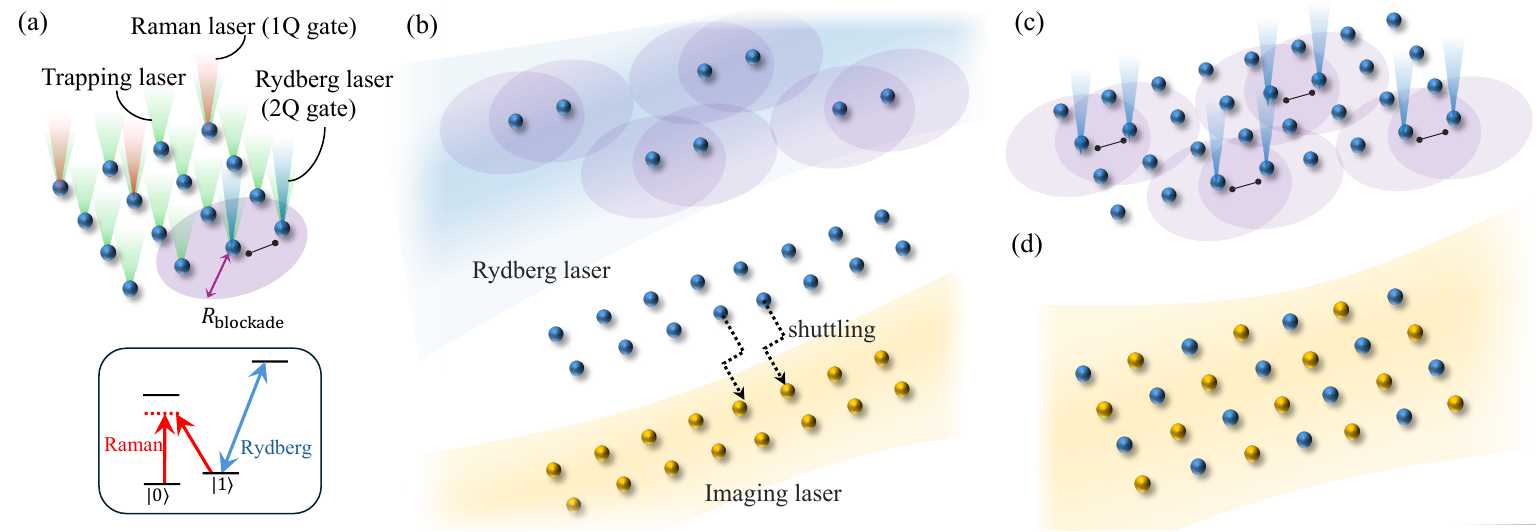}
    \caption{Two approaches for neutral-atom qubit operations.
    (a) Overview of neutral-atom qubit platform, with focused trapping lasers and control laser beams used to perform gate operations.
    (b) Zoned layout, where long-distance atom shuttling is used to switch between two-qubit gate operation and measurements.
    (c-d) Selective operation with addressed control lasers.
    (c) Selective Rydberg gates can be parallelized for a large-scale atom array, with careful arrangement of the pairs to avoid crosstalk. 
    (d) Crosstalk-free measurement, where target atoms (yellow) are being fluoresced by the excitation laser with negligible crosstalk to the other atoms (blue).
    }
    \label{fig:atoms}
\end{figure*}

In this section, we review the operation of neutral-atom qubits as a representative example of a reconfigurable qubit platform, as well as the preliminaries of surface-code FTQC, highlighting the current operational capabilities and state-of-the-art theories, and propose a scalable implementation that serves as the basis of our overall framework.
In Sec.~\ref{sec:intro-atom}, we provide an overview of the neutral-atom array platform for implementing the FTQC, focusing on two prominent physical implementations employed: the zoned and the selective gate approach.
In Sec.~\ref{sec:intro-surface}, we also review surface-code FTQC, in particular, the recent development of the transversal-gate fault-tolerant protocol.
In Sec.~\ref{sec:selective-rydberg}, we propose a scalable implementation of surface-code quantum error correction (QEC) with neutral-atom systems by combining the selective Rydberg gates, selective measurements, and atom shuttling, thereby harnessing the benefits of both zoned and stationary atom array approaches. 

\subsection{Neutral atom platform}\label{sec:intro-atom}

Two approaches for operating physical gates with the neutral atom array are illustrated in Fig.~\ref{fig:atoms}.
In most realizations, single-qubit gates can be operated selectively by addressed control lasers~\cite{jenkins2022ytterbium, levine2022dispersive, lis2023midcircuit}, with fidelity exceeding 99.9\,\% at a gate speed of less than a microsecond.
For other types of operations, a common approach is the zoned layout \cite{bluvstein2022quantum}, where physically separated regions are used for different physical operations.
Two-qubit controlled phase gates, including $CZ$, are performed using the Rydberg blockade effect by exciting atoms from one of the qubit states to highly excited Rydberg states within a typical timescale of several hundred nanoseconds.
The fidelity is currently above 99.5\,\%~\cite{evered2023high-fidelity,peper2025spectroscopy,muniz2024high-fidelity,tsai2025benchmarking,finkelstein2024universal} and is expected to reach 99.9\,\% with better excitation protocols and improved laser systems~\cite{evered2023high-fidelity, jiang2023sensitivity,peper2025spectroscopy,tsai2025benchmarking}.
Combined with selective single-qubit gates, physical CNOT gates can be operated by a sequence $(I\otimes H) CZ (I\otimes H)$ applied to a pair of qubits to interact.
As the Rydberg blockade occurs for atoms within a fixed distance $R_\text{blockade}$, which typically ranges from several $\mu$m to tens of $\mu$m for a chosen Rydberg state used for the interaction, programmable and arbitrary pairwise interactions are typically designed via an atom rearrangement sequence.
In this process, pairs of atoms that need to interact are placed within a certain radius when a global Rydberg excitation laser is applied, inside a \textit{Rydberg zone}~\cite{bluvstein2022quantum,evered2023high-fidelity}.
Qubit measurements are also performed in a separate region, where the fluorescence signal from the atoms is collected to perform state-selective measurements.
While this approach was essential for the recent demonstration of high-fidelity gates, atom rearrangement is expected to be a dominant time cost for the implementation of FTQC.
In particular, for a large system with thousands of qubits, relying on long-distance shuttling for repeated low-level operations necessary in FTQC will ultimately limit the operational speed; more fundamentally, the resulting error accumulation is expected to become even more significant as the system size increases.
Even for trapped atoms with second-scale coherence time, this error accumulation can be substantial, making it an unavoidable, dominant error contribution to the overhead of FTQC\@.

An alternative way to achieve flexible operations has also emerged recently \cite{graham2022multi-qubit, radnaev2025universal, poole2025architecture}, where tightly focused laser beams are used to selectively excite atoms to the Rydberg state, as illustrated in Fig.~\ref{fig:atoms}(c). 
For example, suppose that atoms are arranged in a square grid of lattice spacing $L$, with Rydberg blockade radius $R_\textbf{blockade}$ slightly above $L$. 
By applying the Rydberg lasers to the neighboring sites and not the other, it is thus possible to operate selective $CZ$ gates between the two atoms.
In contrast, a direct application of a global Rydberg laser induces many-body dynamics that are not suitable for digital quantum computation.
For larger arrays, it is possible to perform pairwise interaction in parallel with constant time overhead:
simultaneous application of $CZ$ gates is possible by separating the pairs by, for example, two lattice sites, as illustrated in Fig.~\ref{fig:atoms}(c).
Switching of the addressing pattern is nearly a $\mu$s \cite{radnaev2025universal} with prospects for even faster operation.
This makes it possible to perform the sequential operation of different sets of selective gates within just $\mu$s order or less \cite{radnaev2025universal}, in contrast to the atom shuttling, which is at least a few tens of $\mu$s even for a single shuttling over a $\mu$m-order distance \cite{lam2021demonstration}.

The selective-gate approach is compatible with shuttling-free measurement operations.
This can be achieved using dual-species arrays~\cite{singh2022dual-element, singh2023mid-circuit, anand2024dual-species, nakamura2024dualisotope}, single-species arrays of atoms with metastable qubit states such as Ytterbium and Strontium~\cite{chen2022analyzing,lis2023midcircuit}, and single-species arrays of alkaline atoms such as Rubidium, where addressed shifting lasers \cite{norcia2023midcircuit,hu2024site} or Zeeman-state shelving \cite{graham2023midcircuit} are employed.
All the above approaches exploit either natural or engineered differences in resonant frequencies of atomic transitions between the measured and other atoms so that only desired atoms interact with the imaging laser beams used to excite atomic transitions, leaving the rest largely unaffected.
Small crosstalk of the selective mid-circuit measurement was demonstrated in recent works such as Refs.~\cite{singh2023mid-circuit,lis2023midcircuit}.

\begin{figure*}
    \centering
    \includegraphics[width=0.99\textwidth]{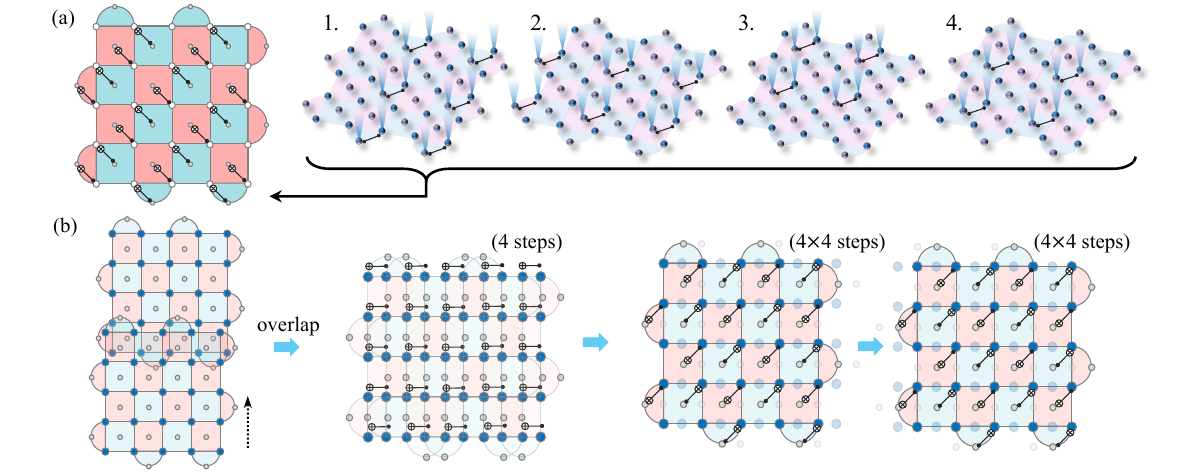}
    \caption{Surface codes with operations implemented by selective Rydberg gates and atom shuttling. 
    (a) Syndrome extraction can be performed by fixed $4\times4$ steps of selective Rydberg $CZ$ gates and selective $H$ gates, where the time cost is limited by the addressing pattern switching time of $\approx 1 \mu$s \cite{radnaev2025universal}. One of four steps of the syndrome extraction circuit for rotated surface code is shown.
    (b) Performing transversal CNOT gate and syndrome extraction after the gate with qubit shuttling and selective Rydberg gate. 
    Atom shuttling is used to move a code block to overlap with another qubit array that forms a logical qubit, creating a dense array with sufficient lattice constant to perform selective Rydberg gate operations. 
    After the atom shuttling, both the transversal gate and the subsequent syndrome extraction can be performed in constant time, as illustrated.
    }
    \label{fig:basic}
\end{figure*}

\subsection{Surface code and logical gates}\label{sec:intro-surface}
Quantum error-correcting codes are essential for implementing large-scale quantum computing while suppressing errors sufficiently to obtain reliable computational outcomes.
A quantum code with $n$ physical qubits, $k$ logical qubits, and distance $d$ is denoted by a $[[n,k,d]]$ code.
A particularly attractive choice is the $[[n,k=1,d=\sqrt{n}]]$ rotated surface code~\cite{bravyi1998quantumcodeslatticeboundary}, which offers a circuit-level noise threshold of around $0.1$--$1$\,\% implemented with two-dimensional local connectivity~\cite{yoshida2025concatenatecodessavequbits,vuillot2019code}, and reduced overhead from the original unrotated version~\cite{orourke2024comparepairrotatedvs}.
Using a polynomial-time decoder based on minimum-weight perfect matching (MWPM), the threshold theorem for fault-tolerant surface-code protocols shows that as long as the physical error rate is kept below a certain threshold, the logical error rate can be exponentially suppressed by increasing the code distance $d$~\cite{10.1063/1.1499754,PhysRevLett.109.180502,10.5555/2685179.2685184,tamiya2024polylogtimeconstantspaceoverheadfaulttolerantquantum,takada2025doublypolylogtimeoverheadfaulttolerantquantumcomputation}.
Neutral atom implementation of the surface codes admits straightforward incorporation of rich real-time information for improving the performance of quantum error correction, such as erasure information~\cite{wu2022erasure, sahay2023high-threshold}, error biases~\cite{tuckett2019tailoring}, and soft information~\cite{majaniemi2025reducing, pattison2021improved}, with circuit-level noise threshold surpassing far beyond $1$\,\% and even approaching $10$\,\% for certain cases~\cite{wu2022erasure,sahay2023high-threshold, sahay2025error}, thus making it a viable option for FTQC\@.

For implementing logical operations, hardware with strict two-dimensional local connectivity would limit the choices of implementation methods, with lattice surgery being the most viable method~\cite{horsman2012surface, litinski2019game}.
In contrast, for neutral atoms with atom shuttling capability, it is possible to realize the transversal implementation of logical gates~\cite{bluvstein2024logical, sahay2025error, cain2024correlated}, introducing an interesting prospect for FTQC\@.
In particular, with transversal gates, the number of measurement rounds in syndrome extraction for the surface code may be as small as one per logical gate, using the technique of correlated decoding and algorithmic fault tolerance with polynomial-time decoding~\cite{zhou2024algorithmic, cain2024correlated, cain2025fast}.
Logical Pauli gates are not to be implemented directly, and instead be tracked in software by updating the Pauli frame.
The logical Hadamard ($H$) gate is implemented by a transversal physical gate followed by the rotation of the code block.
The logical phase ($S$) gate is implementable using an auxiliary state $\ket{Y}\coloneqq\frac{1}{\sqrt{2}}(\ket{0}+i\ket{1})$ catalytically~\cite{fowler2012surface,gidney2017slightly}; alternatively, it is possible to implement $S$ using gate teleportation with $\ket{Y}$, while a fold-transversal implementation of $S$ using atom shuttling and physical $CZ$ gates is also a practical  option~\cite{moussa2016transversal,chen2024transversal}. 

Non-Clifford ($T$) gates are implemented by gate teleportation, where the low-error logical magic state $\ket{T} \coloneqq\frac{1}{\sqrt{2}}(\ket{0}+e^{i\pi/4}\ket{1})$ must be prepared as an auxiliary resource state.
For any target error rate $\epsilon$, a logical $\ket{T}$ state with its logical error rate below $\epsilon$ can be obtained by state injection~\cite{li2015magic,Litinski2019magicstate} and magic state distillation~\cite{bravyi2005universal,bravyi2012magic-state,wills2024constantoverheadmagicstatedistillation}.
In a finite regime of $\epsilon$, efficient magic state preparation protocols, with in-place error suppression using only a single code block, are also actively developed~\cite{hirano2024leveraging, gidney2024magic, chen2025efficient}; however, their scalability to the regime of arbitrarily small target error rates is still an open question due to their growing overhead for small target error rates while this is required for large-scale FTQC\@.
In practice, combining such efficient protocols with magic state distillation is expected to be a viable option to seamlessly interpolate these two regimes.

The above gate set, combined with fault-tolerant preparation of logical $\ket{0}$ and $\ket{+}\coloneqq\frac{1}{\sqrt{2}}(\ket{0}+\ket{1})$ and transversal $X$ and $Z$ measurements~\cite{fowler2012surface}, achieves the implementation of universal quantum computation.

\subsection{Surface-code QEC with selective Rydberg gates}\label{sec:selective-rydberg}

The local gate operations described in Sec.~\ref{sec:intro-atom} (Fig.~\ref{fig:atoms}(c-d)) can be used to implement the rotated surface code by arranging the atoms in a square grid.
Syndrome extraction can be performed by scheduling the data and syndrome qubit interactions as illustrated in Fig.~\ref{fig:basic} (a), where each of four interactions in the surface-code syndrome extraction circuit decomposes into four addressed Rydberg laser reconfigurations~\cite{tomita2014low-distance}.
Selective measurement of syndrome qubits completes the syndrome extraction, either by direct imaging in a dual-species array or by a selective shelving operation in a single-species array~\cite{chen2022analyzing, lis2023midcircuit,graham2023midcircuit}. 

This approach considerably simplifies the system design by eliminating the need for spatially separated zones, thereby fully avoiding the costly atom shuttling among zones.
As illustrated in Fig.~\ref{fig:basic}(b), a transversal CNOT gate can also be performed by adding the qubit shuttling capability, where two atom arrays are overlapped by shuttling to implement transversal atom manipulations with selective Rydberg and Raman lasers.
Syndrome extraction after the gate operation is possible with the arrays overlapped; again with a constant time cost by using addressed Rydberg gate operations and selective imaging.
Notably, this approach makes it possible to perform syndrome extraction flexibly, anywhere in the computing module.
On the other hand, the zone-based approach requires shuttling of syndrome qubits for syndrome extraction within and among spatially separated zones, generally incurring larger time costs for larger systems and offering limited flexibility to perform syndrome extraction during data- or syndrome-qubit shuttling.
Given this critical difference in scalability, we propose implementing neutral-atom qubit operations without dedicated zones and treating each surface-code block and associated syndrome qubits as a standalone unit that can operate on its own, without the need to shuttle qubits across different zones.
In the next section, we will define this standalone unit as a \textit{cell}, which serves as a fundamental building block of our framework.

\section{Building blocks of the game}
\label{sec:logical-cell}

\begin{figure*}[t]
    \centering
    \includegraphics[width=6in]{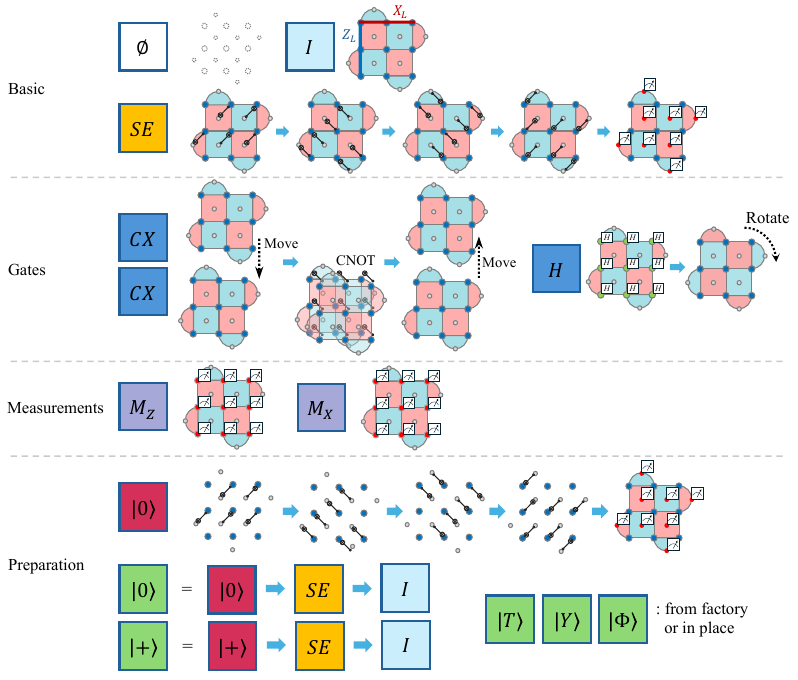}
    \caption{
    Modes of cells.
    A cell is associated with a single code block of a quantum error-correcting code and auxiliary qubits for syndrome extraction, with the rotated surface code being our particular choice in this paper. 
    At any given time, a cell is in one of the available modes.
    Each mode means that the cell is involved in implementing the corresponding fault-tolerant operation of the code, with more detailed definitions given in the main text.
    The empty mode $\emptyset$ is illustrated by a white box, meaning an empty placeholder. 
    The idle mode $I$ is illustrated by a light blue box, representing a surface code block with $Z$ and $X$ stabilizer generators depicted as pink and light blue squares, respectively, and logical $X$ and $Z$ operators shown as red and blue thick lines, respectively. 
    The $SE$ mode is illustrated by the yellow box, implementing a sequence of selective CNOT gates and transversal measurements.
    The gate modes are illustrated by the blue boxes, where $CX$ involves moving a cell to its neighboring cell, performing transversal CNOT gates, and then moving the cell back, and $H$ consists of transversal $H$ gates, followed by a $\pi/2$ rotation of the code block.
    The measurement modes $M_{Z(X)}$ are illustrated by purple boxes, with red data qubits measured transversally. 
    The preparation modes are illustrated by the green boxes, representing a fault-tolerant preparation of the logical qubit in state $\ket{0}$, $\ket{+}$, $\ket{T}$, $\ket{Y}$, or $\ket{\Phi}$.
    For the case of $\ket{0}$ and $\ket{+}$, the fault-tolerant preparation is performed by a noisy state preparation (the red boxes), followed by syndrome extraction.
    For the feasibility of our analysis, this red noisy preparation is treated as a tentative mode that may appear only as part of the overall fault-tolerant state preparation process during the preparation modes.
    The fault-tolerant preparation of $\ket{T}$, $\ket{Y}$, and $\ket{\Phi}$ is conducted within a factory, i.e., a dedicated zone to perform state injection followed by distillation protocols, which may be combined with efficient in-place state-preparation protocols such as magic state cultivation, as discussed later in Fig.~\ref{fig:factory}.
    }
    \label{fig:mode}
\end{figure*}

In this section, we introduce the notion of \textit{cells}, which serves as the fundamental building blocks in our framework for FTQC formulated as the transversal surface-code game.
Roughly speaking, a cell represents the minimal unit of surface code used in our FTQC architecture, capable of transitioning between available \textit{modes} to implement their logical operations.
In Sec.~\ref{sec:logical_cells}, we define the cells together with their modes.
In Sec.~\ref{sec:modes}, we explain the functionality of the modes in terms of fault-tolerant operations on the surface code.
In the remainder of this section, we focus on formally defining a cell by modeling its function in surface-code FTQC with transversal gates, while deferring implementation details for neutral-atom arrays to Sec.~\ref{sec:operating}.

\subsection{Cells}
\label{sec:logical_cells}
A cell in our framework consists of $n$ physical data qubits arranged as a single code block of the $[[n,k=1,d=\sqrt{n}]]$ rotated surface code, together with $n-1$ auxiliary syndrome qubits required for syndrome extraction (SE).
In addition, each cell may include a surrounding spatial region that serves as a workspace, enabling efficient execution of certain logical gate operations such as cell rotation (see also Sec.~\ref{sec:operating} for details on physical implementation).

At any given time, a cell is characterized by its current \textit{mode} of operation, as shown in Fig.~\ref{fig:mode}.
The default modes $\emptyset$ (empty) and $I$ (idle) denote modes with no active operation, used for the cell without or with an encoded logical qubit, respectively.
The mode \textit{SE} (syndrome extraction) is used while the cell is conducting the syndrome extraction for quantum error correction.
While performing logical gates, the cells are in the gate modes of \textit{CX} (CNOT gate) and $H$ (Hadamard gate), respectively.
During the measurements, the cells are in the measurement modes: $M_Z$ ($Z$-basis measurement), and $M_X$ ($X$-basis measurement), respectively.
While preparing the encoded logical qubit in a logical state, the cell is in the preparation modes, namely, $\ket{0}$ ($\ket{0}$-state preparation), $\ket{+}$ ($\ket{+}$-state preparation), $\ket{T}$ ($\ket{T}$-state preparation), $\ket{Y}$ ($\ket{Y}$-state preparation), and $\ket{\Phi}$ ($\ket{\Phi}$-state, i.e., maximally entangled remote qubit pair, preparation), respectively.
Note that the mode for preparing remote $\ket{\Phi}$ states is relevant for implementing large-scale FTQC in a multimodule approach, but is optional for small-scale FTQC using only a single module without the need for interconnection between separate modules.

The default modes $\emptyset$ and $I$ may operate in arbitrarily short time $\tau\geq 0$ as these are allowed to immediately transition to another mode.
The other modes are characterized by their respective operation time $\tau_{SE}$, $\tau_{CX}$, $\tau_{H}$, $\tau_{M_Z}$, $\tau_{M_X}$, $\tau_{\ket{0}}$, $\tau_{\ket{+}}$, $\tau_{\ket{T}}$, $\tau_{\ket{Y}}$, and $\tau_{\ket{\Phi}}$, which are given by some positive real numbers.
Also, the preparation modes are accompanied by associated space-time costs depending on the target logical states to be prepared.
We discuss concrete estimates of these numbers for neutral-atom qubits, in Sec.~\ref{sec:operating}.

\subsection{Modes of cells}
\label{sec:modes}
We now associate each mode to a fault-tolerant operation on logical qubits of the surface code, as illustrated in Fig.~\ref{fig:mode}.
As we will see in Sec.~\ref{sec:ruleset}, any quantum circuit can be simulated with a sufficient number of cells with logical operations supported by these modes.

The mode $\emptyset$ corresponds to the case where the cell is empty. 
This mode is used as a placeholder where a logical qubit can be initialized or moved.

The mode $I$ is the idle mode, where the cell holds a logical qubit awaiting its next operation.
Due to the accumulation of idling errors, $I$ may need to transition to the mode for syndrome extraction regularly.

The mode $SE$ is for the syndrome extraction, representing a single round of measuring stabilizer generators performed by $4\times4$ steps of selective Rydberg $CZ$ gates and selective $H$ gates, followed by selective measurement of syndrome qubits (see also Fig.~\ref{fig:basic}).
For the $[[n, k = 1, d = \sqrt{n}]]$ surface code, measurements in a syndrome extraction are conventionally repeated for $d$ rounds; when multiple rounds of syndrome extraction are required, the cell remains in the $SE$ modes multiple times until all measurement rounds are completed.
On the other hand, the more recent protocols for transversal gates with algorithmic fault tolerance~\cite{zhou2024algorithmic, cain2024correlated, cain2025fast} may require only a single measurement round for syndrome extraction.
Our framework is designed to be compatible with both approaches.

The mode $CX$ is to implement the logical CNOT gate between a pair of cells that are next to each other. 
This is performed by moving one of the neighboring cells to the other, performing transversal CNOT gates, followed by moving the cell back.

The mode $H$ performs the logical Hadamard gate operation. 
For the rotated surface code, physical Hadamard gates are applied to all data qubits in the cell, followed by the rotation of the qubit array to reverse the change of the stabilizer generators and logical operators of the code block.

The mode $M_Z$ ($M_X$) performs the $Z$($X$)-basis measurement of the logical qubit, respectively.
These are performed by transversal $Z$($X$)-basis measurements of data qubits.

The modes $\ket{0}$ and $\ket{+}$ are used during the fault-tolerant preparation of the logical qubit of a cell in $\ket{0}$ and $\ket{+}$, respectively.
This is possible by a noisy preparation of each data physical qubit of the cell in $\ket{0}$ or $\ket{+})$, respectively, followed by syndrome extraction.
As discussed above for the $SE$ mode, syndrome extraction in the $[[n, k = 1, d = \sqrt{n}]]$ surface code may require either $d$ measurement rounds in the conventional approach or a single measurement round in protocols employing transversal gates with algorithmic fault tolerance~\cite{zhou2024algorithmic, cain2024correlated, cain2025fast}.
Again, our framework is compatible with both approaches.

The modes $\ket{T}$, $\ket{Y}$, and $\ket{\Phi}$ represent the fault-tolerant preparation of $\ket{T}$, $\ket{Y}$, and remote $\ket{\Phi}$ across different modules, respectively.
In our framework, $T$ and $S$ gates can be implemented via gate teleportation, utilizing the above operations assisted by the preparation of auxiliary states $\ket{T}$ and $\ket{Y}$, respectively.
The preparation of $\ket{T}$ can be implemented by first performing a noisy state preparation via state injection, followed by magic state distillation; alternatively, the preparation may be implemented in place via magic state cultivation~\cite{gidney2024magic, chen2025efficient} if this suffices for the required target error rate, or by combining both as needed.
For the injection and distillation process, we assign a dedicated zone composed of cells, which we call a factory (see also Sec.\ref{sec:factory} for more details).
When using factories, a cell adjacent to the factory can be set to the preparation mode and will remain in this mode until the factory completes the fault-tolerant preparation of the target state and outputs it into the cell.
In the case of in-place preparation, any cell may enter the preparation mode and will remain in this mode until the preparation process is complete.
The preparation of $\ket{Y}$ can also be achieved either via a factory~\cite{PhysRevA.74.032319,litinski2019game} or in place~\cite{Gidney2024inplaceaccessto, chen2024transversal} (see also Sec.~\ref{sec:factory} for more details).
In large-scale FTQC realized with multimodule architectures, cells can be additionally equipped to support remote gates and quantum teleportation across different modules, using interfaces connected by inter-module links~\cite{sunami2025scalable}.
To enable this, the cell may optionally support a mode for preparing remote $\ket{\Phi}$ states; that is, high-fidelity remote logical Bell states required for gate or state teleportation.
The preparation of remote $\ket{\Phi}$ is achieved via a factory~\cite{matsumoto2003conversion}.
A general and efficient procedure for designing the factories within our framework will be presented in Sec.~\ref{sec:factory}.

\section{Transversal surface-code game}\label{sec:ruleset}

\begin{figure*}[t]
    \centering
    \includegraphics[width=1\linewidth]{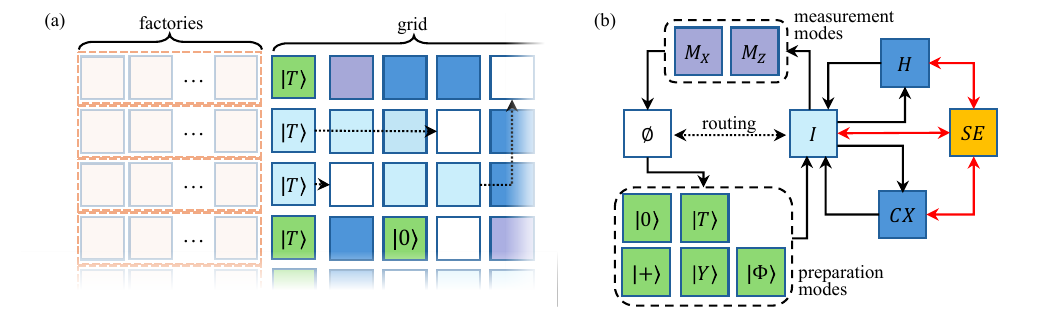}
    \caption{
    A layout of the grid of cells and factories, and the mode-transition rule set for the transversal surface-code game.
    (a) In this game, the given original circuit is implemented on a two-dimensional grid of cells, assisted by factories for fault-tolerant preparation of auxiliary states required for gate teleportation.
    Black dotted arrows indicate the routing used to move the logical states of the surface code on the grid.
    (b) The rule set for transition between modes of the cells on the grid, as detailed by rules~\ref{rl:init}--\ref{rl:meas} in the main text. 
    In this game, the player must ensure that accumulated error parameters do not exceed the error budget determined by the threshold of fault-tolerant protocols, by appropriately transitioning to the $SE$ mode for syndrome extraction (red arrows) according to rules~\ref{ec:1}–\ref{ec:3}.
    }
    \label{fig:rule}
\end{figure*}

Building on the cells and their modes, we now introduce our transversal surface-code game by presenting its goal and rule set.
In Sec.~\ref{sec:game-rule}, we introduce the rules of the transversal surface-code game.
Then, in Sec.~\ref{sec:game-universal}, we supplement the rule set by showing how combinations of these rules can be used to implement various gates, to demonstrate that universal quantum computation can be achieved within this framework.

\subsection{Game setup and rule set}\label{sec:game-rule}

In the transversal surface-code game, a player is initially given a fixed target error $\epsilon>0$ for FTQC and an original quantum circuit expressed in terms of the Clifford+$T$ gate set.
The goal is to implement the original circuit fault-tolerantly using the surface code and, as in the conventional setting of FTQC~\cite{gottesman2010introduction}, guarantee that the computational output is sampled from a probability distribution that is within $\epsilon$ in total variation distance of the original circuit's output distribution.
To this end, the player designs a two-dimensional layout for a grid of the cells along with the factories, as shown in Fig.~\ref{fig:rule}(a).
The game is then played by implementing the original circuit through a (parallel) sequence of mode transitions for the cells on the grid, which must follow a set of mode-transition rules introduced below.
The fault tolerance is ensured by the rule set of the game, where the most crucial requirement is that the syndrome extraction should be performed as soon as the accumulated physical error rate approaches a certain error budget $p_\text{budget}$, that is, the maximum allowable accumulation of physical errors, set below the threshold of the fault-tolerant protocol.
Space-time efficiency can be enhanced by playing the game in an optimized way, for example through optimizing the grid layout, cell routing, and semantics-preserving circuit synthesis.
With the mode-transition sequence specified, the player may adapt the code distance of the surface code for each cell as needed to achieve the target error $\epsilon$ required by the goal.

At the initial time $t=0$, all the cells are set in the $\emptyset$ mode, from which the cells are allowed to implement logical operations according to the \textit{mode-transition} rule set, as described below (see also Fig.~\ref{fig:rule}(b)):
\begin{enumerate}[label=R\arabic*]
    \item\label{rl:init} Logical state preparation ($\emptyset \rightarrow$~preparation mode~$\rightarrow I$): An empty cell  $\emptyset$ may transition to one of the preparation modes, $\ket{0}$,~$\ket{+}$,~$\ket{T}$,~$\ket{Y}$, and~$\ket{\Phi}$, to wait for the logical qubit to be fault-tolerantly prepared in the desired logical state.
    For in-place preparation, any empty cell on the grid may transition to the relevant preparation mode.
    However, some of these states---namely, $\ket{T}$, $\ket{Y}$, and $\ket{\Phi}$---may be prepared in factories. For these states, the transition from $\emptyset$ to the relevant preparation mode is allowed only for empty cells on the grid that are adjacent to the corresponding factories.
    Each mode may require a preparation time $\tau_{\ket{0}}$,~$\tau_{\ket{+}}$,~$\tau_{\ket{T}}$,~$\tau_{\ket{Y}}$, and~$\tau_{\ket{\Phi}}$, respectively.
    After this initialization, the cell transitions to the idle mode $I$\@.
    \item\label{rl:wait} Wait ($I \leftrightarrow SE$): An idling cell may transition to the $SE$ mode for syndrome extraction.
    Conventionally, for the code distance $d$, stabilizer generator measurements in the $SE$ mode are repeated $d$ times for each syndrome extraction.
    With the algorithmic fault tolerance, a single round of measurement in the $SE$ mode may suffice for each syndrome extraction~\cite{cain2024correlated,zhou2024algorithmic, cain2025fast}.
    The time cost of syndrome extraction is denoted by $\tau_\text{SE}$.
    After the syndrome extraction, the cell returns to the idle mode $I$.
    \item\label{rl:H} Hadamard gate ($I \leftrightarrow H \leftrightarrow SE$): An idling cell may transition to the $H$ mode to implement a logical $H$ gate, which requires a time $\tau_H$.
   As the code distance increases, $\tau_H$ may also grow, and errors may accumulate during this period. For fault tolerance, the cell in the $H$ mode can transition to the $SE$ mode during the execution, by splitting the operation into multiple sections, and then return to the $H$ mode to complete the remainder of the gate implementation  (see also Sec.~\ref{sec:hadamard}). After the logical gate is applied, the cell returns to the $I$ mode. 
    \item\label{rl:CX} CNOT gate ($I \leftrightarrow CX\leftrightarrow SE$): A pair of adjacent idling cells can transition to the $CX$ mode to implement a logical CNOT gate at time cost $\tau_{CX}$.
    Similar to the $H$ mode, as the code distance increases, $\tau_{CX}$ may also grow, and errors may accumulate during this period. For fault tolerance, the cells in the $CX$ mode can transition to the $SE$ mode by splitting the operation into multiple sections, and then return to the $CX$ mode to complete the remainder of the gate implementation. After the logical gate is applied,  the cells return to the $I$ mode.
    \item\label{rl:shuttle} Routing ($(I,\emptyset) \rightarrow (\emptyset,I)$): If the cell is idle $I$, it can exchange positions with an empty cell $\emptyset$, with an associated atom-shuttling time cost $\tau_r$.
    During this routing, the cell remains in the mode $I$, ensuring that it can transition to the $SE$ mode during the routing operation by splitting the routing into multiple steps if needed.
    \item\label{rl:meas} Measurement ($I \rightarrow$~measurement mode~$\rightarrow \emptyset$): An idling cell can transition to the measurement modes $M_X$ or $M_Z$ to perform logical-qubit measurements in $X$ or $Z$ bases, respectively, at time costs $\tau_{M_X}$ and $\tau_{M_Z}$, respectively. After the measurement, the cell transitions to the $\emptyset$ mode, requiring reinitialization for subsequent use.
\end{enumerate}

In this game, the player must ensure that accumulated errors do not exceed the predetermined error budget $p_\text{budget}$.
To achieve this, the player keeps track of an accumulated error parameter for each cell and executes the decoding according to the following rules.

\begin{enumerate}[label=E\arabic*]
\item \label{ec:1}Initially, the accumulated error parameter in the empty mode $\emptyset$ is set to zero.
The preparation modes, i.e., $\ket{0}$, $\ket{+}$, $\ket{T}$, $\ket{Y}$, and $\ket{\Phi}$, have associated error parameters $p_{\ket{0}}$, $p_{\ket{+}}$, $p_{\ket{T}}$, $p_{\ket{Y}}$, and $p_{\ket{\Phi}}$, respectively.
These error parameters represent the logical error rates in the fault-tolerant preparation of each state, all assumed to be smaller than $p_\text{budget}$.
After completing the mode, its error parameter is added to the accumulated error parameter of the cell.

\item\label{ec:2} For modes $m \in \{I, H, CX\}$, the accumulated error parameter is given by a monotonically increasing function $p_m(t)$, where $t$ is the time spent in the mode.
As the cell remains in mode $m$, the accumulated error parameter increases by $p_m(t)$.
In the case of $CX$ modes, in addition to the error accumulation $p_m(t)$ for the cell itself, error propagation from its interacting cell is also taken into account.
When the cell transitions from mode $m$ to another mode at time $\tau_m$, the accumulated error parameter has increased by $p_m(\tau_m)$, which is carried over to the next mode.

\item\label{ec:3} When the accumulated error parameter reaches the error budget $p_\text{budget}$, the cell is transitioned to the $SE$ mode, even if it is in the middle of implementing a logical gate operation.
Based on measurement outcomes from the $SE$ mode, the player concurrently performs quantum error correction by running the classical decoder for the surface code, which resets the accumulated error parameter.
Also in the measurement modes $M_X$ and $M_Z$, the logical measurement outcomes are determined by the decoder, resetting the accumulated error parameter similar to the $SE$ mode.
\end{enumerate}

In this way, the rules of the game ensure that syndrome extraction is always performed before accumulated errors approach the error budget set below the threshold, thereby guaranteeing the feasibility of FTQC\@.

\subsection{Implementing universal quantum computation within the game}\label{sec:game-universal}

\begin{figure*}[t]
    \centering
    \includegraphics[width=1\linewidth]{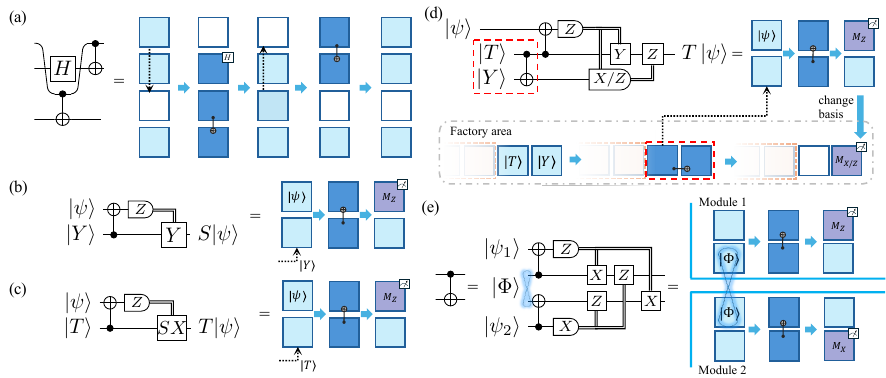}
    \caption{
    Implementation of gates required for universal quantum computation via mode-transition rules.
    (a)
    Demonstration of implementing long-range CNOT gates using routing and the $CX$ mode for nearest-neighbor CNOT gates. A CNOT gate between the first and third input qubits in the circuit on the left can be implemented by routing the cell accordingly.
    (b) Implementation of $S$ gates via gate teleportation assisted by the auxiliary state $\ket{Y}$. The cell prepared in $\ket{Y}$ is routed next to the target cell, followed by $CX$ and measurement $M_Z$ of the target cell. The resulting Pauli correction is handled by updating the Pauli frame. 
    (c) Conventional implementation of $T$ gates via gate teleportation assisted by the auxiliary state $\ket{T}$. The cell prepared in $\ket{T}$ is routed next to the target cell,
    followed by $CX$ and measurement $M_Z$ of the target cell. This implementation requires an $S$ correction, which in turn may require routing a cell prepared in $\ket{Y}$ as described in (b). 
    (d) Optimized implementation of $T$ gates that we propose. In this implementation, $\ket{T}$ and $\ket{Y}$ are pre-interacted at the time of their preparation. Then, as in the conventional gate teleportation, the cell initially prepared in $\ket{T}$ is routed next to the target cell, followed by $CX$ and measurement $M_Z$ of the target cell.
    The cell initially prepared in $\ket{Y}$ may remain fixed.
    Conditioned on the measurement outcome of the target cell, the cell initially prepared in $\ket{Y}$ is measured either in $M_X$ or $M_Z$, realizing the delayed-choice $S$ correction with improved time efficiency.
    The resulting Pauli correction is handled by updating the Pauli frame.
    (e) Implementation of remote CNOT gates between two logical qubits in different modules.
    The cells prepared in $\ket{\Phi}$ are routed next to the target cells in each module, followed by $CX$ and measurements $M_Z$ and $M_X$ of the target cells, respectively. The resulting Pauli correction is handled by updating the Pauli frame.
    }
    \label{fig:gate-ops}
\end{figure*}

We demonstrate that the composition of the above mode transitions allows us to implement universal quantum computation.
The following examples illustrate how to combine mode transitions to implement the original circuit, which may include long-range CNOT gates, $S$ gates, and $T$ gates, as well as how to implement interfaces for establishing interconnects between multiple modules.
\begin{enumerate}[label=C\arabic*]
    \setcounter{enumi}{0}
    \item\label{rl:CNOT} Long-range CNOT gate: We can implement a long-range CNOT gate between an arbitrary pair of cells on the grid, by combining the routing and the $CX$ mode for the nearest-neighbor CNOT gate, as shown in Fig.~\ref{fig:gate-ops}(a)\@.  
    \item\label{rl:S}
    The $S$ gate:
    The combination of $CX$, $M_Z$, and $\ket{Y}$ enables gate teleportation to implement the $S$ gate, as shown in Fig.~\ref{fig:gate-ops}(b).
    In this implementation, the auxiliary state $\ket{Y}$ is routed next to the target cell, followed by a $CX$ mode between the routed cell in $\ket{Y}$ and the target cell.
    The target cell is then measured by the $M_Z$ mode, resulting in the $S$ gate applied to the output state up to a Pauli $Y$ correction.
    In our framework, Pauli gates are always tracked by updating the Pauli frame.
    As shown in the figure, we propose to use gate teleportation that outputs the state to the newly prepared $\ket{Y}$ cell.
    Compared to outputting the state to the target cell, this approach has the notable advantage of better robustness against leakage errors in the target cell, since such errors may be detected through the measurement in the teleportation to a freshly prepared auxiliary cell.
    While an alternative implementation of $S$ gates using catalytic $\ket{Y}$ is known~\cite{fowler2012surface, gidney2017slightly}, this method may be less robust to leakage errors and generally incurs a higher time cost than gate teleportation.
    For these reasons, computations on the grid should primarily use the proposed gate-teleportation method, while the catalytic implementation of $S$ can be used for efficiently preparing $\ket{Y}$ in the factories, as discussed in Sec.~\ref{sec:y-factory}.
    \item\label{rl:T}
    The $T$ gate:
    The combination of $CX$, $M_Z$, $M_X$, $\ket{T}$, and $\ket{Y}$ enables gate teleportation for implementing the $T$ gate, as shown in Fig.\ref{fig:gate-ops}(c), with its optimized version shown in Fig.\ref{fig:gate-ops}(d).
    In the conventional gate teleportation shown in Fig.~\ref{fig:gate-ops}(c), the auxiliary state $\ket{T}$ is routed next to the target cell, followed by $CX$ between the routed cell for $\ket{T}$ and the target cell.
    The target cell is then measured in the $M_Z$ mode, resulting in the $T$ gate applied to the output state, up to $S$ and Pauli $X$ correction.
    The $S$ correction is applied using the methods described above with $\ket{Y}$.
    In the optimized version that we propose in Fig.~\ref{fig:gate-ops}(d), rather than routing both $\ket{T}$ for gate teleportation and $\ket{Y}$ for correction, we interact $\ket{T}$ and $\ket{Y}$ via $CX$ in advance upon their preparation.
    Subsequently, the $\ket{T}$ cell is routed next to the target cell, followed by $CX$ between the routed $\ket{T}$ and the target. The $\ket{Y}$ cell may remain fixed.
    The target cell is then measured in the $M_Z$ mode, and conditioned on the measurement outcome, the $\ket{Y}$ cell is measured in either $M_X$ or $M_Z$ to determine whether an $S$ correction is applied.
    A particular advantage of this proposed method is that it eliminates the need to route the $\ket{Y}$ cell, allowing for correcting $S$ by just changing the measurement basis without additional routing.
    This improves the time cost and simplifies the routing schedule for gate teleportation.
    In addition, as with the $S$ gate teleportation, the state is output to the newly prepared $\ket{T}$ cell to ensure robustness against leakage errors.
    \item\label{rl:phi} Inter-module operation: If the preparation mode for remote $\ket{\Phi}$ over different modules is supported, the combination of $CX$, $M_X$, $M_Z$, and $\ket{\Phi}$ enables gate teleportation for implementing the remote CNOT gate between modules, as shown in Fig.\ref{fig:gate-ops}(e), as well as quantum teleportation of states between the modules.
    In this gate teleportation, the auxiliary state $\ket{\Phi}$ is routed next to the target cells in each module, followed by $CX$ between the routed cells in $\ket{\Phi}$ and the target cells.
    The target cells are then measured by $M_Z$ and $M_X$, resulting in the CNOT gate applied to the output state, up to Pauli corrections that are tracked by updating the Pauli frame.
    Similar to above, the state is output to the newly prepared $\ket{\Phi}$ cells to ensure robustness against leakage errors.
\end{enumerate}

Together with the $H$ gate supported by the $H$ mode, the above examples demonstrate that any original circuit in the Clifford+$T$ gate set, composed of $H$, $S$, $T$, and CNOT gates between arbitrary pairs of qubits, can be implemented within our framework.

\section{Transversal-gate-based resource state factory}\label{sec:factory}

\begin{figure*}[pt]
    \centering
    \includegraphics[width=0.87\linewidth]{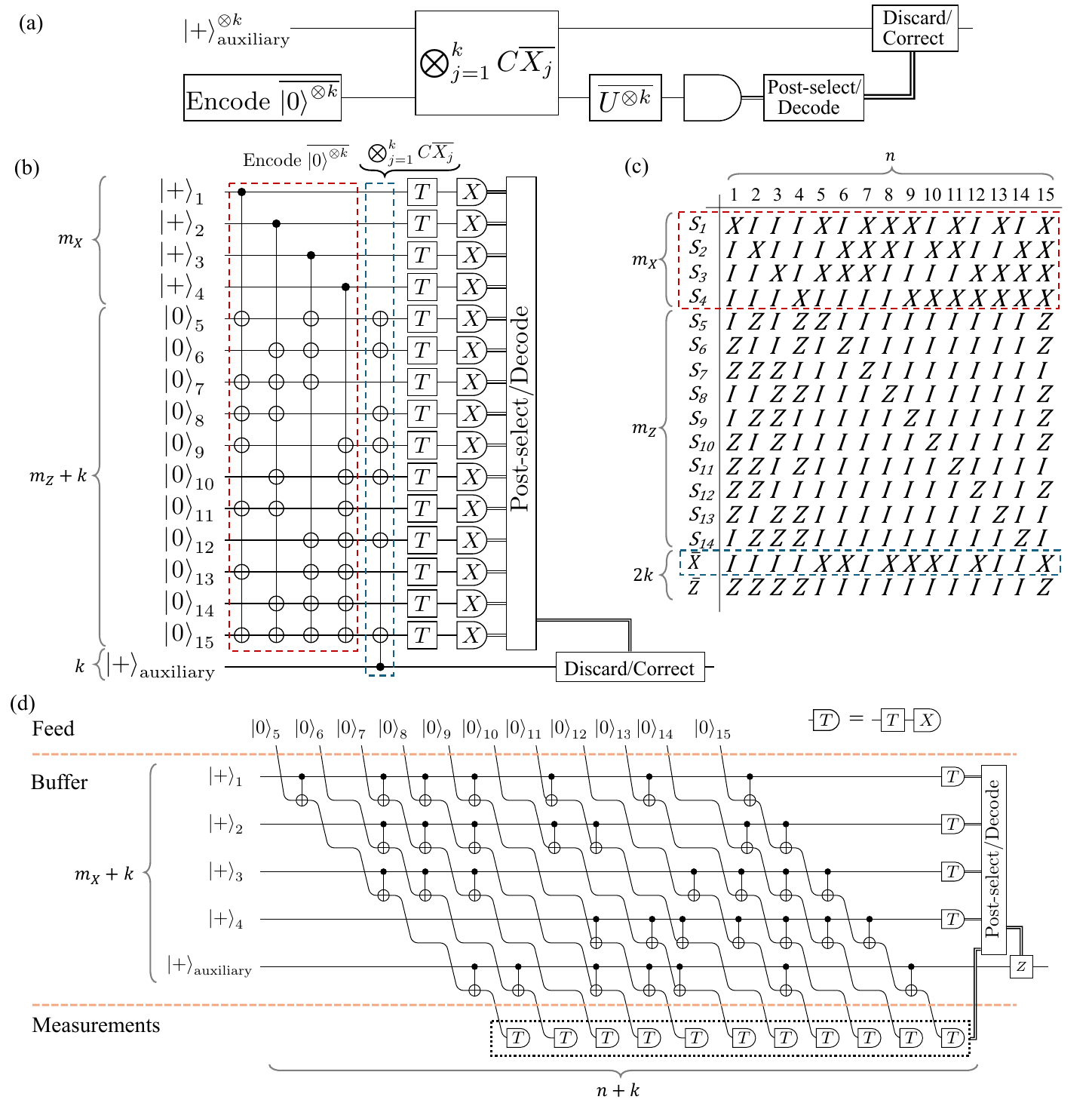}
    \caption{Resource state distillation circuit and an example of magic state distillation using $[[15,1,3]]$ code.
    (a) An outline of resource state distillation based on an $[[n,k,d]]$ CSS code $\mathcal{C}$.
    In this circuit, $k$ auxiliary qubits are initialized in the state $\ket{+}^{\otimes k}$, and the remaining $n$ qubits are encoded into the logical state $\overline{\ket{0}^{\otimes k}}$ of $\mathcal{C}$.
    Then, $k$ controlled-$\overline{X}$ operations, $\bigotimes_{j=1}^{k}C\overline{X_j}$, are applied between the $j$th auxiliary qubit (as control) and the corresponding $j$th logical $X$ operator of $\mathcal{C}$ (as target), thereby generating $k$ maximally entangled pairs between the auxiliary qubits and the logical qubits.
    This is followed by a transversal application of the logical gate $\overline{U^{\otimes k}}$ on $\mathcal{C}$, implemented as a tensor product of noisy gates $\bigotimes_{i=1}^{n} U_i'$ acting on the $n$ physical qubits of the code.
    Finally, all qubits in $\mathcal{C}$ are measured in a suitable basis.
    Given the measurement outcomes, we either perform post-selection for error detection by discarding the output if any error is detected, or apply a decoding algorithm for error correction.
    Up to the correction, this process completes the preparation of a high-fidelity resource state for implementing $\overline{U^{\otimes k}}$ via gate teleportation.
    (b–-c) An example of a magic state distillation circuit for $\ket{T}$ using the $[[15,1,3]]$ code, shown alongside a table that represents the stabilizer generators and logical operators of the code.
     In the table, the first $14$ rows correspond to the $X$ and $Z$ stabilizer generators presented in the standard form of Ref.~\cite{cleve1997efficient}, with each column corresponding to one of the $15$ physical qubits of the code (see main text for details). The final two rows specify the logical $X$ and $Z$ operators of the encoded logical qubit.
     The first stage of the circuit (enclosed in the red dashed box) prepares the logical state $\overline{\ket{0}^{\otimes k}}$ as in Refs.~\cite{cleve1997efficient, gottesman1997stabilizer}: qubits 1–4 are initialized in the $\ket{+}$ state, and multitarget CNOT gates are applied based on the $X$ stabilizer generator pattern specified in the table.
Next, a control-$\overline{X}$ gate (blue dashed box) is applied using multitarget CNOTs, where the targets correspond to the support of the logical $X$ operator in the table.
This is followed by a transversal application of the logical $\overline{T}$ gate, realized as $T^{\otimes 15}$ (up to some Clifford correction).
All $15$ physical qubits of the $[[15,1,3]]$ code are then measured in the $X$ basis, yielding $15$ classical bits for post-selection or decoding.
Up to correction, the resulting state of the auxiliary qubit is a high-fidelity $\ket{T}$ state.
    (d) A pipelined circuit equivalent to that in (b), with $5$ buffer qubits and $11$ pipelined feed qubits moving from top to bottom. See also Fig.~\ref{fig:factory} for efficient implementation of this circuit with cell routing and transversal gates.
    }
    \label{fig:factory-basic}
\end{figure*}

Distillation of resource states, such as $\ket{T}$, $\ket{Y}$, and $\ket{\Phi}$, is crucial for the scalable implementation of FTQC, and within our framework, the space-time cost of preparing these states is expected to be a major factor in the overall system performance, as is generally the case for FTQC\@.
In this section, we introduce a general state factory design for the transversal-gate FTQC protocol and discuss its efficient implementation within our game rules.
First, in Sec.~\ref{sec:css-distillation}, we discuss a general procedure for constructing resource state distillation circuits using $[[n,k,d]]$ quantum codes.
In Sec.~\ref{sec:factory-design}, we outline their implementation within our architecture, following the rules of the game.
For better space-time efficiency, in Sec.~\ref{sec:factory-compact}, we propose their efficient, pipelined implementation and analyze its space-time cost.
Finally, in Sec.~\ref{sec:y-factory}, we further propose an efficient factory design particularly optimized for $\ket{Y}$.
\subsection{Resource state distillation protocols}\label{sec:css-distillation}

State distillation is crucial for generating error-suppressed auxiliary resource states for gate teleportation,
which enables the implementation of a unitary gate $U$ that cannot be directly implemented transversally on the surface code, and hence serves as a fundamental building block in FTQC\@.
Examples of such resource states include $\ket{T}$, $\ket{Y}$, and $\ket{\Phi}$, as shown in Fig.~\ref{fig:gate-ops}.
To prepare such a state via distillation, we concatenate the surface code with another quantum error-correcting code that implements the target unitary $U$ transversally.
Then, after performing state injection on the surface code, this approach enables the (generally noisy and non-fault-tolerant) application of transversal $U$ by gate teleportation, which is then followed by error detection or correction to suppress errors.
We begin by reviewing the notations, followed by a description of the factory circuit construction for the distillation.

Consider an $[[n,k,d]]$ CSS code $\mathcal{C}$ defined by a set of independent $n-k$ stabilizer generators $S = \{S_i\}_{i=1,...,n-k}$.
Among these stabilizer generators, let $m_X$ denote the number of $X$ stabilizer generators, i.e. those given by a tensor product of either identity $I$ or Pauli $X$ operators on $n$ qubits. 
In this section, we refer to the logical qubits encoded in the surface codes simply as qubits, while those encoded in $\mathcal{C}$ are referred to as logical qubits.
The remaining $m_Z = n-k-m_X$ stabilizer generators are the $Z$ stabilizer generators.
Also we write the set of $X$ and $Z$ logical operators of $\mathcal{C}$ for the $k$ logical qubits as $\{\overline{X_j}\}_{j=1,\ldots,k}$ and $\{\overline{Z_j}\}_{j=1,\ldots,k}$, respectively.
A gate $U$ is transversal for $\mathcal{C}$ if a logical gate is implemented by the tensor product of a unitary gate $U'$ on the $n$ qubits comprising the code, $\overline{U^{\otimes k}} = \bigotimes_{i=1}^n U'_i$ where $U_i'$ may be different from the target unitary $U$ in general.
For example, in magic state distillation for preparing $\ket{T}$, one often uses an $[[n,k,d]]$ triorthogonal code that enables the logical $\overline{T^{\otimes k}}$ gate via the transversal application of $T^{\otimes n}$ up to a Clifford correction~\cite{bravyi2012magic-state}. 
In such cases, assuming that the Clifford correction can be performed in a fault-tolerant way, such as gate teleportation with $\ket{Y}$, we focus here on the essential part of distillation: the application of transversal $T^{\otimes n}$ on the $[[n,k,d]]$ code.

We illustrate the general structure of distillation circuits with code $\mathcal{C}$ in Fig.~\ref{fig:factory-basic}(a), along with a concrete example for magic state distillation with $[[n=15,k=1,d=3]]$ code (with $m_X=4$ and $m_Z=10$)~\cite{bravyi2005universal} in Figs.~\ref{fig:factory-basic}(b) and (c). 
Whereas this example distills $\ket{T}$,  the $[[n=7,k=1,d=3]]$ Steane code ($m_X=m_Z=3$) can be used for the distillation of $\ket{Y}$ and $\ket{\Phi}$~\cite{matsumoto2003conversion,PhysRevA.74.032319,litinski2019game}.
The circuit runs on $k+n$ qubits, with the first $k$ auxiliary qubits and the $n$ qubits for encoding $\mathcal{C}$.
First, the latter $n$ qubits are encoded into $k$ logical qubits of $\mathcal{C}$, followed by the generation of maximally entangled states between the first $k$ auxiliary qubits and the $k$ logical qubits encoded in $\mathcal{C}$.
The encoding circuit for a CSS code can be obtained from the ``standard'' form of its stabilizer table, i.e., a tabular representation of the code's stabilizer generators, as shown in Fig.~\ref{fig:factory-basic}(c) for the $[[15,1,3]]$ code.
In this figure, the first through fourteenth rows correspond to the stabilizer generators, with each column representing a physical qubit of the code.
Note that the choice of stabilizer generators is not unique; one can manipulate the matrix to arrive at a specific ``standard form,'' and we here adopt the form defined in Ref.~\cite{cleve1997efficient} by applying Gaussian elimination.
This standardization is useful for constructing the encoding circuit, as we will see below.
We also append $X$ and $Z$ logical operators at the bottom of this table.
In general, this table has $n$ columns and $n+k$ rows, where the first $m_X$ rows correspond to $X$ stabilizer generators, $m_Z$ rows to the $Z$ stabilizer generators, followed by $k$ rows for the $X$ logical operators and $k$ rows for the $Z$ logical operators.
To prepare the logical $\overline{\ket{0}^{\otimes k}}$ state, we use the $m_X$ $X$ stabilizer generators from the table; in particular, we initialize $m_X$ qubits in $\ket{+}$, and then, for each $X$ stabilizer generator $S_i$ with $i\in\{1,...,m_X\}$,  we apply a control-$S_i$ operation targeting the remaining $n-m_X=m_Z+k$ qubits, using the $i$th qubit as the control~\cite{cleve1997efficient}.
These operations can be represented as multitarget $CX$ gates, as shown in Fig.~\ref{fig:factory-basic}(b).
The remaining $m_Z$ $Z$ stabilizer generators are automatically satisfied with this encoding.
To generate $k$ Bell pairs between the $k$ auxiliary qubits and the $k$ encoded logical qubits, we then apply, for each $j\in\{1,\ldots,k\}$, a control-$\overline{X_j}$ operation, using the $j$th auxiliary qubit (prepared in $\ket{+}$) as control.
These can again be represented as multitarget $CX$ gates.
Overall, this part of the circuit for preparing the $k$ Bell pairs has width $n+k$ and depth $m_X+k$, and is illustrated in Fig.~\ref{fig:factory-basic}(a).

After preparing these Bell pairs, the target logical gate $\overline{U^{\otimes k}}$ on $\mathcal{C}$ is implemented transversally by applying noisy $\bigotimes_{i=1}^{n}U_i'$ gates on the $n$ physical qubits of $\mathcal{C}$:
for our example in Fig.~\ref{fig:factory-basic}(b), $U_i'=T$ is applied by noisy $\ket{T}$ state preparation and gate teleportation, thereby enacting $\overline{U^{\otimes k}}=\overline{T}$ on the logical qubit with $k=1$.

Finally, $n$ physical qubits encoding $\mathcal{C}$ are measured in the appropriate basis, yielding a classical bit string $(b_1,\ldots,b_n)\in\{0,1\}^{n}$: for our example of magic state distillation, qubits are measured in the $X$ basis to detect or correct $Z$ errors, yielding $15$ classical bits.
These measurement results can then be used to evaluate the stabilizer generators of $\mathcal{C}$ to perform either error detection or correction:
for $X$ stabilizer generator $S_i$ that acts nontrivially on a set $S_\mathrm{S}^{(i)}$ of qubits, we obtain the parity by evaluating $\sum_{l\in S_\mathrm{S}^{(i)}}b_l\mod 2$, which is $0$ in the absence of errors and may be flipped to $1$ if errors are present. 
Similarly, for a logical Pauli operator $\overline{X_j}$ with $S_{\mathrm{L}}^{(j)}$, the logical measurement result is given by the parity $\sum_{l\in S_{\mathrm{L}}^{(j)}}b_l\mod 2$.
For the case of error detection, if any of the stabilizer generators are measured to be $1$, the output state is discarded; if the error probability for each non-fault-tolerant application of $U_i'$ on the surface code is $p$, and the code distance of $\mathcal{C}$ is $d$, then the error rate of the post-selected output state is $O(p^d)$.
If quantum error correction is used instead, the measurement outcomes of the stabilizer generators are handed to a decoding algorithm, which determines any necessary corrections for the output state and can substantially reduce overhead, especially at large scales~\cite{wills2024constantoverheadmagicstatedistillation}; in this case, if the decoder can correct up to $t$ errors (for example, $d=2t+1$ if $d$ is odd and the decoder corrects up to half the code distance), the output error rate is $O(p^{t+1})$, and the state is produced deterministically.

Finally, in Fig.~\ref{fig:factory-basic}(d), we propose a pipelined circuit that is suitable for transversal implementation.
This circuit is equivalent to that in Fig.~\ref{fig:factory-basic}(b), differing only by temporal and spatial rearrangement of the qubits and wires. 
The auxiliary $k$ qubits and the $m_X$ qubits for the $X$ stabilizer generators in Fig.~\ref{fig:factory-basic}(b) are grouped to form the buffer.
The remaining $n-m_X$ qubits constitute the feed, sequentially initialized and moved into the buffer, where they interact with the buffer qubits via two-qubit CNOT gates following the same pattern as in Fig.~\ref{fig:factory-basic}(b).
After these interactions, each feed qubit is moved into the measurements part of the circuit, undergoes the transversal gate $U_i'$, and is measured as in Fig.~\ref{fig:factory-basic}(b).
The $m_X$ qubits in the buffer also undergo the transversal gate $U_i'$ and are subsequently measured. 
Importantly, the pipelined circuit is not to be implemented by SWAP gates; instead, the movement of the feed qubits is realized through routing in our framework.
As we will see in Sec.~\ref{sec:factory-design}, this approach allows efficient pipelined preparation, one-way movement, and measurement, all with a small space usage of at least $m_X + k$ while the circuit depth is bounded by $n+k$.

\begin{figure*}[t]
    \centering
    \includegraphics[width=0.95\linewidth]{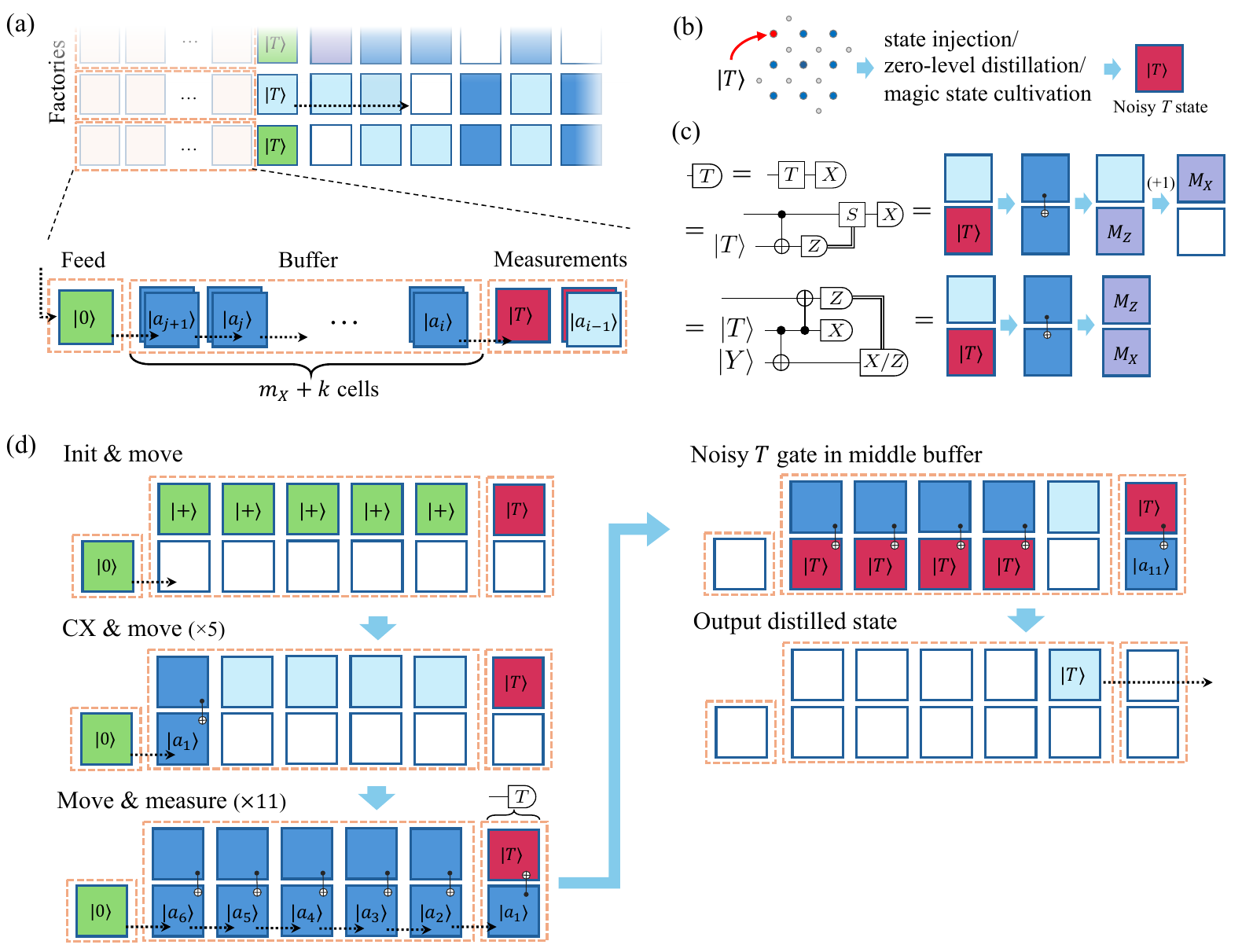}
    \caption{
    Resource state factory using the cells implementing the distillation circuit of Fig.~\ref{fig:factory-basic}(d).
    This design straightforwardly generalizes to state distillation with other target resource states or those based on other CSS codes (see the main text for details).
    (a) A pipelined resource state factory based on the one-way routing of cells. 
    When illustrating overall layout (as in Figs.~\ref{fig:rule} (a) and \ref{fig:processor}), we represent cells for the factory in faint orange colors to abstract out its internal operations (top).
    (b) The noisy state preparation mode. The $\emptyset$ cell turns into a noisy resource state (red) by either state injection, zero-level distillation protocols, or magic state cultivation.
    (c) Noisy gate teleportation and measurement for the $\ket{T}$ distillation circuit execution.
    The circuit in the middle is the gate teleportation based on adaptive $S$ gate execution.
    With the protocol shown in the bottom panel (delayed-choice correction; Pauli corrections are omitted), the time cost reduces by delegating the adaptive operation to the measurement bases of $\ket{Y}$ states.
    (d) Detailed breakdown of the mode transitions in the factory.
    The space-efficient implementation of this protocol is described further in Fig.~\ref{fig:factory-compact}.
    }
    \label{fig:factory}
\end{figure*}

\subsection{Factories for transversal-gate architectures}\label{sec:factory-design}
Having constructed the distillation circuits, we present a resource state factory design using the rules of the transversal surface-code game.
Our overall design is shown in Fig.~\ref{fig:factory}(a), consisting of the feed area for fault-tolerant initialization of a cell in $\overline{\ket{0}}$, a buffer for interacting $m_X+k$ stationary cells with each feed cell routed sequentially, and the measurement area for performing the noisy gate teleportation and measurements.
This design applies to a general class of distillation protocols presented in Sec.~\ref{sec:css-distillation} while the example in Fig.~\ref{fig:factory}(a) implements the pipelined circuit in Fig.~\ref{fig:factory-basic}(d) for magic state distillation~\cite{bravyi2012magic-state}.

In Fig.~\ref{fig:factory}(a), the feed on the left repeatedly supplies cells initialized in $\overline{\ket{0}}$, which are sequentially routed through the buffer in a pipelined manner.
Within the buffer, transversal $CX$ gates are applied between the routed feed cell (representing the feed qubit in the circuit) and the stationary buffer cells (representing the $m_X+k$ buffer qubits in the circuit), according to the structure of the pipelined distillation circuit.
The $U_i'$ gate is implemented by noisy gate teleportation followed by the measurement of the cell.
The noisy gate teleportation is assisted by noisy state preparation (Fig.~\ref{fig:factory}(b)) with state injection~\cite{li2015magic,Litinski2019magicstate,Lao22magic, gidney2023cleaner, jacoby2025magic}, zero-level distillation techniques~\cite {hirano2024leveraging}, or magic state cultivation~\cite{gidney2024magic, chen2025efficient}.
With this noisy state preparation, Fig.~\ref{fig:factory}(c) shows how to implement the gate teleportation and measurement of the cell; in the teleportation of $T$ gates, similar to Fig.~\ref{fig:gate-ops}(d), we can adopt delayed-choice implementation rather than conventional adaptive $S$ corrections.

As discussed in Sec.~\ref{sec:css-distillation}, the present distillation factory design is not restricted to the magic state distillation but is generally applicable to other resource states.
For a target unitary $U$ that is transversely implementable in an $[[n,k,d]]$ CSS code with $m_X$ $X$ stabilizer generators and $m_Z$ $Z$ stabilizer generators, the middle buffer length is determined by parameter $m_X+k$, while the round of routing is $(m_Z+k)+(m_X+k)=n+k$.

\begin{figure}[t]
    \centering
    \includegraphics[width=0.99\linewidth]{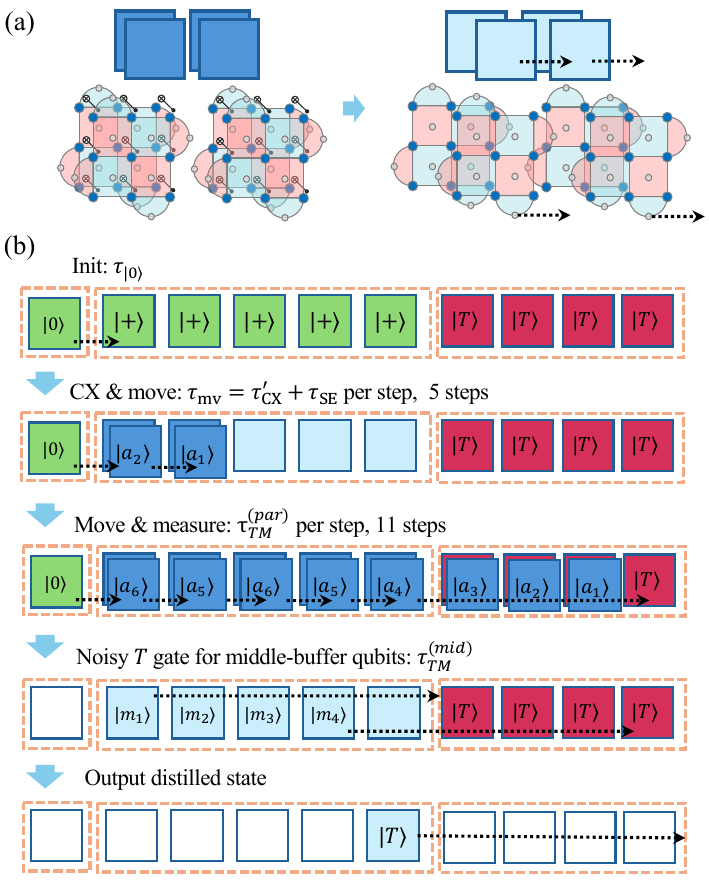}
    \caption{
    Our optimized, pipelined design of the factory with a special treatment of pipelining and parallelization. 
    (a) Optimized routing of the overlapped cells in the factory. CNOT gates, measurements, and one-way routing in the factory are performed in this optimized manner without returning the cells to their original positions.
    (b) Implementation of the factory in Fig.~\ref{fig:factory}(d) using this optimized routing. 
    }
    \label{fig:factory-compact}
\end{figure}

\subsection{Efficient, pipelined factories}
\label{sec:factory-compact}
Based on this factory design, we propose a further optimized design of factories, using pipelining and parallelization. 
The pipelined routing allows parallel execution while maintaining small spatial usage, making it particularly well-suited to our architecture; in particular, this parallelized routing approach is highly efficient for the neutral-atom platforms, as we discuss later in Sec.~\ref{sec:factory-time}.
In our optimization, we adjust the size of the measurement area to reduce the time cost for $\ket{T}$ state preparation and gate teleportation across multiple cells.
Furthermore, we introduce a special treatment of the cell's modes within the factory to enable space-efficient factory design, without altering the hardware requirements or degrading operational speed; specifically, for the predetermined and pipelined routing of cells in the factory, we allow these cells to remain spatially overlapped throughout the distillation process, as illustrated in Fig. \ref{fig:factory-compact}(a).
In this case, both $CX$ and $SE$ can be performed without returning the cells to their original positions, in contrast to the rule in Fig.~\ref{fig:mode} requiring such a return.
This design enables the factory to operate within a single row, as shown in Fig. \ref{fig:factory-compact}(b), without incurring additional operational overhead compared to the two-row design illustrated in Fig.~\ref{fig:factory}(d), resulting in only $m_X+k=5$ cells in the buffer area for the 15-to-1 magic state distillation protocol.

Here we briefly outline the runtime of the factory for magic state distillation while delegating a more detailed analysis to Sec.~\ref{sec:factory-time}.
The time cost within the factory is dominated by the pipelined operation: initialization in the feed $\tau_{\ket{0}}$, routing in the buffer taking $\tau_\text{mv}$ per step, as well as the operations in the measurement area taking $\tau_{TM}^{(par)}$, summarized in Fig.~\ref{fig:factory-compact}(b).
The preparation of $\ket{0}$ takes $\tau_{\ket{0}} \approx 2\tau_{SE}$.
The pipelined routing cost between the CNOT gate is $\tau_{CX}' = \tau_{CX}/2$, where the factor-of-two improvement over $\tau_{CX}$ arises from the fact that there is no need to bring the cell back to the adjacent location, unlike the native $CX$ mode.
A single step of routing and interaction in the middle buffer thus takes $\tau_\text{mv} = \tau_{CX}' + \tau_{SE}$.

The time cost of operations in the measurement area is determined by the cost  $\tau_{\ket{T}}'$ of noisy $\ket{T}$ preparation, and times of gate teleportation, $S$ correction, and measurements.
Noisy $\ket{T}$ state initialization time $\tau_{\ket{T}}'$ is highly dependent on the protocol employed, with state injection completing within several syndrome extraction cycles~\cite{li2015magic, Lao22magic}, as well as zero-level operations and cultivation requiring many retries due to the low success probability~\cite{hirano2024leveraging, gidney2024magic, chen2025efficient}.
With the delayed-choice protocol (Fig.~\ref{fig:factory}(c), bottom), $S$-gate correction in the measurement area can be replaced by adaptive measurement of $\ket{Y}$ resource state, resulting in time-efficient operation, albeit additional routing time for $\ket{Y}$ resource.
It is possible to mitigate the time cost of noisy $T$ state initialization and $\ket{Y}$ state interaction by using multiple cells in the measurement area for parallelized $\ket{T}$ generation.
With the number of the cells available in the measurement area being sufficiently large, state preparation time cost becomes negligible, resulting in $\tau_{TM}^{(par)} \approx \tau_{CX}' + \tau_M$, on par with the gate and routing time in the middle buffer for $\tau_{SE} \approx \tau_M$, and achieving fully pipelined fault-tolerant resource state preparation.

Overall, in the case of $15$-to-$1$ magic state distillation protocol, the time cost for a single factory trial is $\tau_\text{factory} = 5\tau_\text{mv} + 11\tau_{TM}^{(par)} + \tau_{TM}^{(mid)}$, where the first term is required for the first feed cell to reach the measurement area, and the last term $ \tau_{TM}^{(mid)}$ is the $T$-gate teleportation and measurement of the cells in the buffer, which is slightly longer than $\tau_{TM}^{(par)}$ due to the longer routing distance from the buffer to the measurement area.
For the case of post-selection, the average $\ket{T}$-state generation time $\tau_{\ket{T}}$ is expected to be slightly longer than $\tau_\text{factory}$ because of finite discard probability; however, the difference is expected to be small.
We analyze realistic costs for the case of limited space in the measurement buffer, in Sec.~\ref{sec:factory-time}.

\begin{figure}[ttt]
    \centering
    \includegraphics[width=0.99\linewidth]{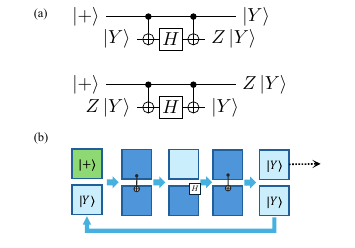}
    \caption{
    A factory for preparing $\ket{Y}$ using the catalytic implementation of $S$ gates.
    (a) Circuits to implement $S$ gates  by catalytically using $\ket{Y}$. Once a single copy of $\ket{Y}$ is fault-tolerantly prepared, for example via distillation, it is possible to catalytically generate multiple copies of $\ket{Y}$ fault-tolerantly~\cite{gidney2017slightly}.
    (b) The mode transition in the two-cell $\ket{Y}$ factory, using a single copy of $\ket{Y}$ catalytically. The $Z$ correction is not explicitly shown in the figure since this can be handled by updating the Pauli frame.
    }
    \label{fig:Y-factory}
\end{figure}

\subsection{Optimized factories for $\ket{Y}$ states}\label{sec:y-factory}

Fault-tolerant preparation of $\ket{Y}$ can be achieved via a noisy state preparation followed by distillation~\cite{fowler2012surface}; alternatively, methods based on lattice surgery~\cite{Gidney2024inplaceaccessto} may be used, yet note that it remains unclear whether these lattice-surgery-based approaches are compatible with techniques that reduce the measurement rounds to one using algorithmic fault tolerance.
By contrast, the transversal-gate-based preparation methods proposed here are compatible with the algorithmic fault tolerance.

Once a sufficient number of $\ket{Y}$ states have been fault-tolerantly prepared, it becomes possible to employ a more efficient $\ket{Y}$-state factory based on a catalytic implementation of $S$ gates~\cite{gidney2017slightly};
this is achieved by fault-tolerant preparation of $\ket{+}$ states followed by applying $S$ gates using $\ket{Y}$ catalytically.
This completes with only one extra cell, with three logical gates as shown in Fig.~\ref{fig:Y-factory}.
As such, for large-scale computations, the space-time cost of $\ket{Y}$ state is dominated only by the catalytic $\ket{Y}$ factory with two cell spaces, which is substantially smaller than the $\ket{T}$ state distillation cost.

\section{Neutral-atom array implementation}\label{sec:operating}

To aid the resource estimation of FTQC with neutral atoms, here we analyze the typical time costs of the modes of the cells.
Routing and the $CX$ mode are discussed in Sec.~\ref{sec:cnot}, measurement modes in Sec.~\ref{sec:meas-mode}, the $SE$ mode in Sec.~\ref{sec:se}, and the $H$ mode is analyzed in Sec.~\ref{sec:hadamard}.
Composite instructions are also discussed, with $S$ and $T$ gates in Sec.~\ref{sec:tel-gate} and the preparation mode $\ket{T}$ via the $\ket{T}$ factory evaluated in Sec.~\ref{sec:factory-time}.

\subsection{Routing and CNOT gate}\label{sec:cnot}
\begin{figure*}[t]
    \centering
    \includegraphics[width=0.99\textwidth]{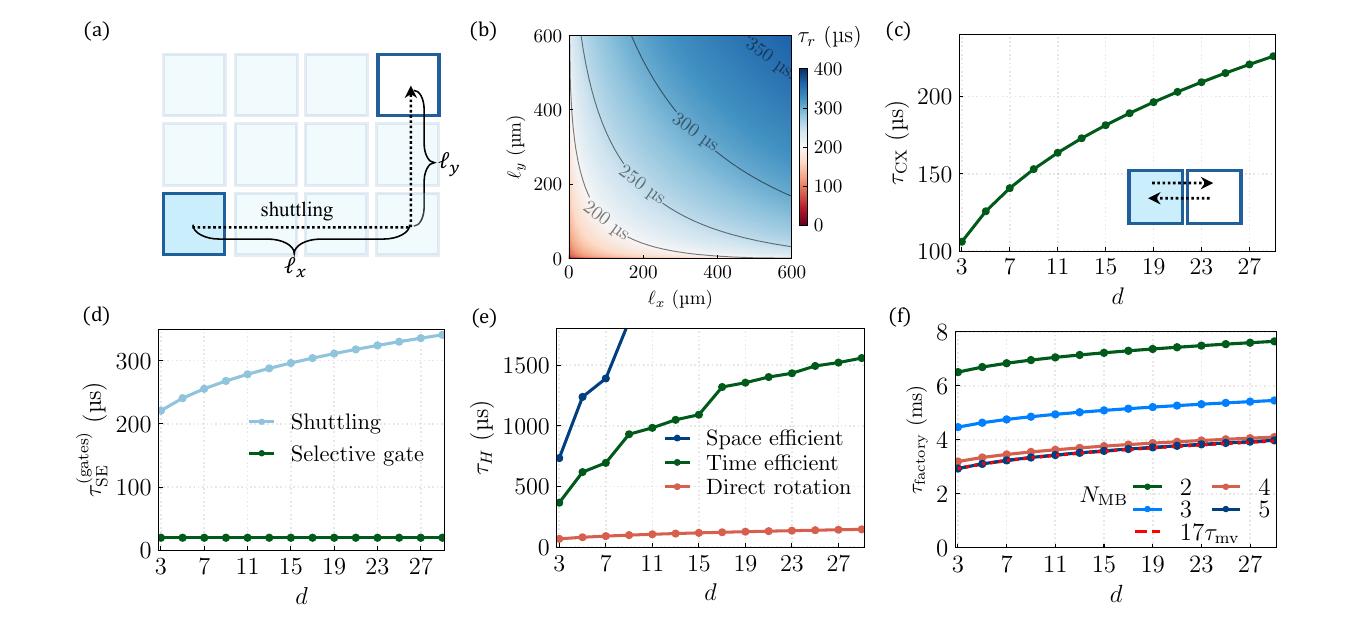}
    \caption{Time cost for implementing logical operations on a neutral atom platform.
    (a) A schematic of atom shuttling. To avoid hitting the atoms in other atoms arranged in a two-dimensional square grid, the two-dimensional shuttling occurs in two steps:
    first a horizontal shuttling by $l_x$, followed by a vertical shuttling by $l_y$. 
    (b) Cell routing time $\tau_r$ via two-dimensional atom shuttling. The time $\tau_r$ is not isotropic, since diagonal movements require shuttling atoms sequentially along two axes, requiring a longer time than horizontal or vertical movements.
    (c) Time cost for $CX$ modes, depending on the code distance of the rotated surface code. This is given by the two steps of one-dimensional atom shuttling over $d\times L$ distance, where $L$ is the lattice constant of the data qubit array.
    As discussed in Sec.~\ref{sec:ruleset}, once the resource state is ready, $\tau_S$ and $\tau_T$ have the same time cost as $\tau_{CX}$.
    (d) Syndrome extraction time $\tau_\mathrm{SE}^{(\mathrm{gates})}$, where the qubit measurement time omitted from the plot. 
    The shuttling model accounts for the time required to move the syndrome qubits inside the block before the SE, as well as outside the code block after the SE, as well as the shuttling times within the block.
    The selective gate refers to the $SE$ protocol proposed in Fig.~\ref{fig:basic}(a).
    (e) Time cost $\tau_H$ of implementing a logical Hadamard gate by rotating the atom array. 
    Blue and green points with connecting lines are the time required to perform divide-and-conquer atom rearrangements of Ref.~\cite{chen2024transversal}, while red points are the minimum time required for direct rotation of the atom array based on Ref.~\cite{hwang2024fast}.
    (f) Runtime $\tau_\mathrm{factory}$ for the single trial of 15-to-1 $\ket{T}$ factory, for qubit measurement time 100 $\mu$s (see the main text). 
    The time cost $\tau_\mathrm{factory}$ is plotted for different numbers  $N_\mathrm{MB}$ of cells in the measurement area. Ideal pipelined time cost (red dashed line) is achieved with $N_\mathrm{MB}=5$.
    }
    \label{fig:times}
\end{figure*}

Routing of the cells in the grid relies on atom shuttling realized by two-dimensional acousto-optic deflectors (AODs). 
A large time cost is required to move the atoms without exciting the motional states via a careful shuttling trajectory.
A common choice is called the constant jerk, which was demonstrated to show relatively fast shuttling without significant decoherence~\cite{bluvstein2022quantum}.
An improved protocol was recently demonstrated using shortcuts to adiabaticity (STA)~\cite{hwang2024fast, Cicali2024statransportoptimization}; this protocol is similar to the minimum-jerk trajectory recently introduced in Ref.~\cite{chinnarasu2025variational}.
The formula for the time cost of shuttling in these protocols is summarized in Appendix~\ref{app:STA}.
Based on this formula, we plot the time cost to route a cell over a trajectory shown in Fig.~\ref{fig:times}(b).

Based on this time cost, following the description in Sec.~\ref{sec:logical-cell}, we consider the time cost for the $CX$ mode as that of the round-trip shuttling over a single cell length.
Then, using the same formula, we plot the time cost for $CX$ in Fig.~\ref{fig:times}(c) for a fixed lattice constant $L=5~ \mu$m, a commonly employed value for recent experimental demonstrations~\cite{radnaev2025universal,bedalov2024fault-tolerant}, while the resulting performance does not strongly depend on this value.

\subsection{Measurements}\label{sec:meas-mode}
Qubit measurement times $\tau_{M_X}$ and $\tau_{M_Z}$ for neutral-atom qubits are now approaching several hundreds of $\mu$s, with recent demonstrations of high-fidelity, non-destructive measurement in the 160 -- 250 $\mu$s range~\cite{shea2020submillisecond, chow2023high-fidelity}.
For destructive measurement, a recent experiment reported high-fidelity measurement over 20 $\mu$s \cite{senoo2025highfidelity,ma2023high-fidelity} after which the qubits are lost from the trap.
Non-destructive measurements were also reported using the low-dark-count-rate single-photon detector \cite{shea2020submillisecond} and optical cavities \cite{deist2022mid-circuit, Grinkemeyer2025cavityerrordetection} with a time cost of several tens of $\mu$s, thus rapid advancements in measurement fidelity and time are expected.
In our estimation, we assume $\tau_{M_{X/Z}} \approx 100~\mu$s, based on a predicted near-term performance analyzed in Appendix~\ref{app:local}\@.

\subsection{Syndrome extraction}\label{sec:se}
The movement-free implementation of SE, illustrated in Fig.~\ref{fig:basic}, requires 16 steps of addressing-beam AOD reconfigurations and Rydberg gates, as well as single-qubit Hadamard gates.
Addressed-laser reconfiguration time of nearly $1~ \mu$s is already demonstrated \cite{radnaev2025universal}, with Rydberg gate and single-qubit fast Raman gates on the order of $100$ ns \cite{jenkins2022ytterbium}.
Based on these values, the total time cost of $SE$ is around $\tau_\text{SE}^\text{(gates)} \approx 20~\mu$s with potential for further reduction~\cite{poole2025architecture}.
In Fig.~\ref{fig:times}(d), we compare this with the time cost of a shuttling-based $SE$ protocol, which consists of the shuttling time within the code block to perform four physical $CZ$ gates, as well as the time cost to move the syndrome qubits in and out of the cell.
Due to the code-distance-dependent array size,
the time cost of shuttling-based SE scales with distance,
in contrast to the constant time for local selective-gate SE\@.
With measurement time for syndrome qubits $\tau_\text{SE}^\text{(meas)}\approx \tau_{M_{X/Z}}$, the total time cost of SE is $\tau_\text{SE} = \tau_\text{SE}^\text{(gates)} + \tau_\text{SE}^\text{(meas)}$.
In particular, under the above assumption $\tau_{M_{X/Z}} \approx 100~\mu$s, we have $\tau_\text{SE}\approx 120~\mu$s.

\subsection{Hadamard gate}\label{sec:hadamard}
A logical Hadamard gate can be applied by performing Hadamard gates on all physical qubits, followed by a $\pi/2$ rotation of the code block to correct for the change of $X$ and $Z$ (see also Fig.~\ref{fig:mode}). 
While the rotation of a two-dimensional array of atoms is not natively supported by the AOD-generated tweezer array, one could use two sets of horizontally and diagonally aligned AODs to implement reflections along the horizontal and diagonal lattice, realizing the rotation by the number of steps logarithmic to the code distance~\cite{chen2024transversal}.
However, the logarithmically growing number of shuttling steps involved results in a growing time cost, as shown
in Fig.~\ref{fig:times}(e).
Additionally, the divide-and-conquer rearrangement of Refs.~\cite{chen2024transversal} requires additional space of around 1.5 times the size of the code block; a space-efficient protocol without additional space usage will result in a factor of 2 increase in the time cost.
The most critical issue with this approach is that, although the time cost increases at large scales and leads to greater decoherence of atoms, it is impossible to transition to SE during the intermediate steps.
As a result, this method fundamentally violates the assumption of the threshold theorem, which requires syndromes to be extracted frequently enough to keep the error accumulation between SEs below a constant threshold; therefore, this approach is also incompatible with the rules of our transversal surface-code game.

We here adopt an alternative implementation that addresses this issue.
In particular, it was recently demonstrated~\cite{hwang2024fast} that the shuttling of tweezer-trapped atoms with a circular trajectory is possible with comparative performance to one-dimensional shuttling, which suggests a prospect for direct rotation of cells, for example, by the use of rotating lattices~\cite{williams2008dynamic} or dynamically generated moving tweezer arrays based on a spatial light modulator (SLM)~\cite{kim2016situ,lin2024ai-enabled,knottnerus2025parallel}.
The particular advantage of this method is that the $SE$ can be performed during the rotation, compatible with the rules of the game.
In Fig.~\ref{fig:times}(e), we plot the estimated time cost for the direct array rotation, based on the theoretical analysis in Ref.~\cite{hwang2024fast} for STA-based trajectory with expected motional heating kept below a single quanta; the achievable time cost is expected to be orders of magnitude smaller than that of AOD-based sorting, thus strongly motivating experimental demonstrations.

\subsection{Gate teleportation for $S$ and $T$ gates}\label{sec:tel-gate}
On rotated surface codes, one could apply $S$ gates fold-transversally to a mid-cycle state during the $SE$~\cite{chen2024transversal}.
This protocol requires atom shuttling inside the cell to apply fold-transversal gates.
The time cost for this protocol is again code-size-dependent (Fig.~\ref{fig:ftrS}), and it is challenging to perform $SE$ during the fold-transversal implementation.

As such, we choose gate teleportation as a native method for $S$ gates in our game (see also Fig.~\ref{fig:gate-ops}), for which we argue that the required auxiliary state $\ket{Y}$ can be efficiently prepared (see also Fig.~\ref{fig:Y-factory}).
Once the required auxiliary state is prepared, the gate time cost is equivalent to $\tau_\mathrm{CX}$, since the logical measurement of the target qubit does not affect the subsequent operations, and the measurement result is only required for Pauli frame updates in software.

Similarly, with the delayed-choice protocol shown in Fig.~\ref{fig:gate-ops}(d), the relevant $T$ gate execution time $\tau_T$ is the same as the $CX$ gate $\tau_\mathrm{CX}$.
Finite decoding time may lock up $\ket{Y}$ for the adaptive measurement, but not the computation on the cell; the target cell that is teleported does not need to wait for the completion of both the first and the second adaptive measurements of the delayed-choice $T$ gate, before starting its next operation. This is because the measurement outcomes of these measurements only update the Pauli frame being tracked by classical software.

\subsection{State preparation in factories}\label{sec:factory-time}

Among the time costs of the preparation modes $\tau_{\ket{0}}$,~$\tau_{\ket{+}}$,~$\tau_{\ket{T}}$,~$\tau_{\ket{Y}}$, and~$\tau_{\ket{\Phi}}$, the time cost of $\tau_{\ket{T}}$ is typically dominant, which we analyze here.
In Sec.~\ref{sec:factory}, we described an efficient, pipelined implementation of the factories for magic state distillation.
As discussed in Sec.~\ref{sec:factory-compact}, the time cost is $\tau_\text{factory} = 5\tau_\text{mv} + 11\tau_{TM}^{(par)} + \tau_{TM}^{(mid)}$, which simplifies to $\tau_\text{factory} \approx 17 \tau_\text{mv} = 17 (\tau_{CX}' + \tau_{SE})$ for $\tau_M \approx \tau_{SE}$ with a sufficiently large number $N_\mathrm{MB}$ of measurement area cells. 
This ideal time cost is realized for $\tau_\text{mv}$ exceeding $\tau_\mathrm{TY}'/N_\mathrm{MB}$, where we define $\tau_\mathrm{TY}'$ as the time to prepare the noisy $\ket{T}$ state and entangle it with the $\ket{Y}$ resource state. 

The noisy $\ket{T}$ state preparation time $\tau_{\ket{T}}'$ strongly depends on the target error rate for the distilled $\ket{T}$ states.
In the state injection schemes of Refs.~\cite{li2015magic,Lao22magic}, with a physical error rate of $10^{-3}$, the injected $T$ states can have an error rate slightly below the physical error rate with moderate post-selection and time cost of several syndrome extraction rounds.
With this performance for injected $\ket{T}$ states, the leading-order error rate of the factory output is around $10^{-8}$. 
For erasure qubits, as appropriate for some neutral-atom species, an efficient post-selection scheme was proposed to achieve a lower injected $T$-state error rate~\cite{jacoby2025magic}.
This allows $10^{-4}$ injected state error rates, given a total physical error rate of $10^{-3}$ and erasure conversion rates above 90\,\% \cite{wu2022erasure, jacoby2025magic}, where the injection time cost is only marginally increased, and the acceptance probability remains high. 
This reduces the error rate of distilled $\ket{T}$ states to $10^{-10}$ or below.
For better performance, state injection can be replaced by magic state cultivation~\cite{gidney2024magic, chen2025efficient} although it may incur a larger time cost $\tau_{\ket{T}}'$ due to low success probability.

In Fig.~\ref{fig:times} (f), we plot the factory single-trial runtime for finite $N_\text{MB}$, assuming a magic-state-injection time cost of 4$\tau_{SE}$ under the assumption $\tau_{SE} \approx 120~\mu$s.
For these parameters, the factory runtime reduces to $17 \tau_{mv}$ with a measurement area size of $N_\text{MB}=5$, while the $N_\text{MB}=4$ results in only slightly slower operation.

\section{Application to resource estimation for FTQC}\label{sec:example}
For a given Clifford+$T$ circuit, our framework now allows the estimation of circuit execution times for a given module configuration and parameters.
As an example, we consider the case of 100 logical qubits executing Clifford+$T$ circuits with $10^8$ $T$ count, with a target error rate of $10^{-10}$.
This setting was used for the resource estimation in Ref.~\cite{litinski2019game}, where the time cost for executing the circuit with superconducting qubits operating lattice surgery was estimated; 
for $10^{-4}$ physical gate error rate and employing 15-to-1 $\ket{T}$ factories, a 76,400-qubit module was estimated to perform the above circuit in 2 hours.
The error-correction cycle time of 1 $\mu$s assumed in Ref.~\cite{litinski2019game}, dominated by the measurement time of superconducting qubits, is more than 100 times faster than the neutral-atom counterpart we assume in this estimation.\footnote{For example, a cycle time of 1.1 $\mu$s for superconducting qubits was reported recently~\cite{acharya2025quantum}, while there are novel approaches actively being developed for faster operations with superconducting qubits, such as Refs.~\cite{spring2025fast,rower2025suppressing}, potentially reaching several hundreds of ns cycle time.}
However, as we discuss below, the speed of physical-level operations does not immediately translate into the difference in computational execution time for the same qubit count, as the underlying architecture design and operational capabilities are vastly different.

\begin{figure}[t]
    \centering
    \includegraphics[width=1\linewidth]{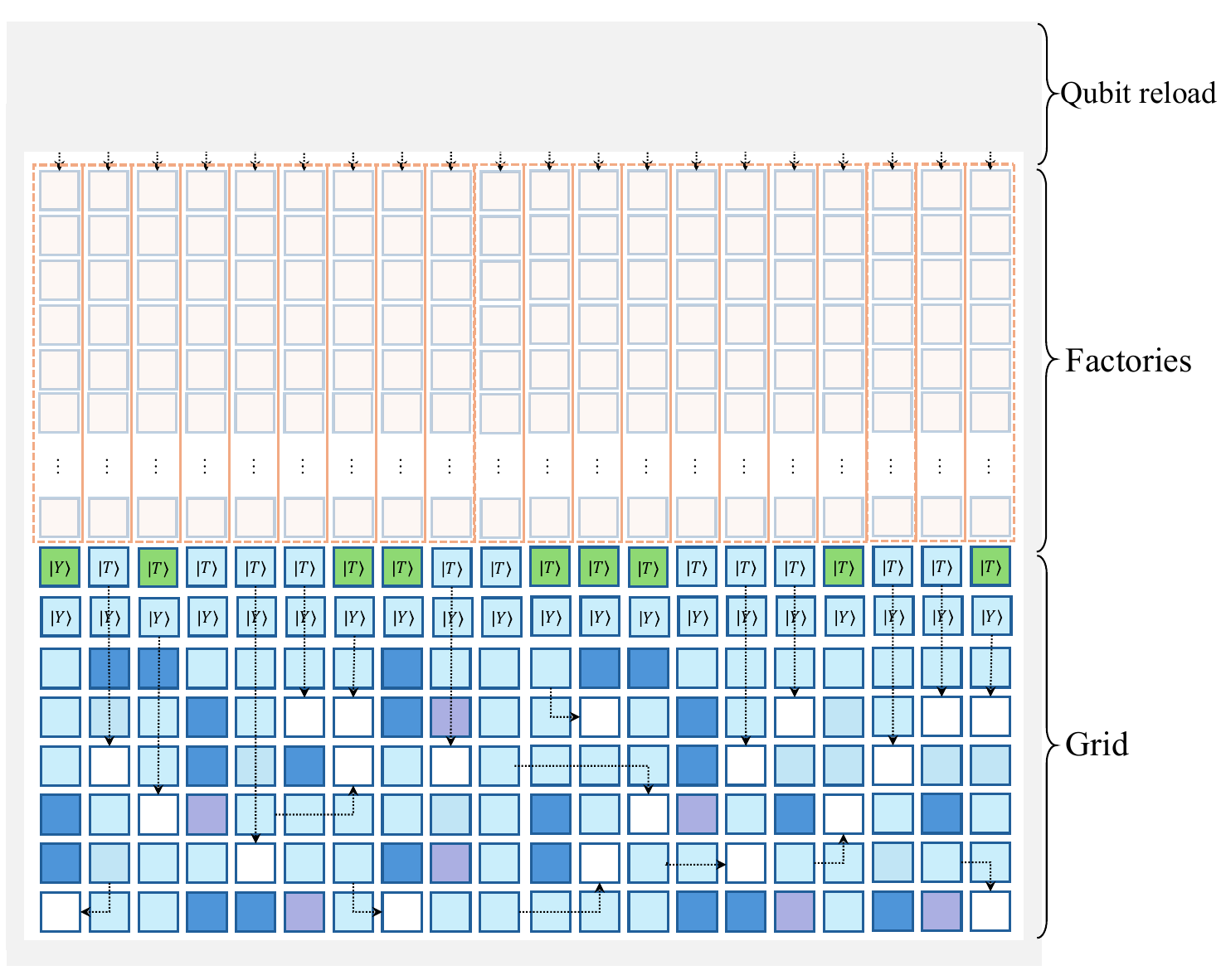}
    \caption{General design of a module for executing Clifford+$T$ circuits.
    Pipelined $\ket{T}$ and $\ket{Y}$ factories are placed at the top of the module, with their output cells situated close to the grid of cells used for implementing quantum computation via the transversal surface-code game. 
    }
    \label{fig:processor}
\end{figure}

\begin{table*}[t]
\centering
\begin{tabular}{ |c|c|c|c| } 
\hline

\multirow{4}{5em}{Input Clifford+$T$ circuit} & width $W$& 100 &\\
& $T$ count per depth & 5 &  \\
& total $T$ count & $10^8$ & \\ 
& depth $D$& $2 \times 10^{7}$ & \\
\hline

\multirow{4}{5em}{QEC params} & target logical error & $10^{-10}$ &  \\
& CZ error $p$ & $10^{-3}$ & \\
& distance $d$ & $9$ & erasure conversion \cite{wu2022erasure,sahay2023high-threshold, sahay2025error} \\ 
\hline

\multirow{3}{5em}{$\ket{T}$ factory} & $\tau_\text{factory}$ & 3 ms & Sec.~\ref{sec:factory-time} \\ 
& $N_\text{factory}$ & 25 & \\
& measurement area $N_\mathrm{MB}$ & 4 & \\
& space cost of a factory (cells) & 13 & \\
& throughput per $\tau_\text{cycle} $ & 5 & $N_\text{factory}\tau_\text{cycle}/\tau_\text{factory}$ \\ \hline

\multirow{5}{5em}{Space costs} & single cell (atoms) & 161 & $N_\text{cell}=2d^2-1$ \\ 
& single $\ket{T}$ factory & 2093 &  $13N_\text{cell}$ \\
& single $\ket{Y}$ factory & 322 &  $2N_\text{cell}$ \\
& $\ket{T}$ factory (total) & 52325 & 25 $\ket{T}$ factories \\
& $\ket{Y}$ factory (total) & 16100 & 50 catalytic $\ket{Y}$ factories \\
& qubits in grid & 16100 &  $100 N_\text{cell}$ \\
\cline{2-3}
& total qubit count  & 76475 & \\
\hline

\multirow{4}{5em}{time costs for EC} & rounds of SE per SE-mode call & $1$ & algorithmic FT \cite{cain2024correlated,cain2025fast, zhou2024algorithmic}\\ 
& selective gates $\tau_\text{SE}^\text{(gates)}$ & 20 $\mu$s & Sec.~\ref{sec:se}\\  
& syndrome-qubit measurements $\tau_{SE}^\text{(meas)}$ & 100 $\mu$s & Sec.~\ref{sec:meas-mode} \\  
& total:  $\tau_{SE} = \tau_{SE}^\text{(meas)}+\tau_{SE}^\text{(gates)}$ & 120 $\mu$s & \\  
\hline

\multirow{8}{5em}{Circuit execution} & logical Hadamard $\tau_H$ & 90 $\mu$s & Sec.~\ref{sec:hadamard} \\ 
& logical  CX $\tau_{CX}$ & 150 $\mu$s & Sec.~\ref{sec:cnot}\\
& measurement $\tau_M$ & 100 $\mu$s & Sec.~\ref{sec:meas-mode} \\ 
& gate round: $\tau_\text{ops} = \max \left[ \tau_H, \tau_\text{M}, \tau_\text{CX}\right]$ & 150 $\mu$s &  $\tau_T, \tau_S \approx \tau_{CX}$,~ Sec.~\ref{sec:tel-gate} \\  
& routing round: $\tau_r$ & 220 $\mu$s &  \\
\cline{2-3}
& cycle time $\tau_\text{cycle} = \tau_r + \tau_\text{ops} + 2\tau_{SE}$ & 610 $\mu$s &  \\  \hline

\end{tabular}
\caption{Parameters used for resource estimation. The $\ket{T}$ factory operates with 3 routing cells active in the buffer area on average.
More than one dedicated $\ket{Y}$ factory is available for each $\ket{T}$ factory for $S$ correction of noisy $T$ gate teleportation inside the factory. 
The total time cost to execute the input circuit is given by $\tau_\text{circuit} = D \times \tau_\text{cycle}\approx 3.4$ hours.
} 
\label{table}
\end{table*}

Parameters for our simplified estimation are summarized in Table \ref{table}, and the module layout is illustrated in Fig.~\ref{fig:processor}. 
Following Ref.~\cite{litinski2019game}, we set the target logical error rate as $10^{-10}$ , for which distance $d=9$ is required with the use of erasure conversion~\cite{wu2022erasure, sahay2025error}, resulting in 16,000 qubits required for hosting $100$ cells.
For non-Clifford gate execution, we add 25 $\ket{T}$ factories on top of the grid which is 25-cell width, with distilled $\ket{T}$ states available for use by routing into the grid.
Together with 50 $\ket{Y}$ factories, the necessary physical qubit count increases to 76,500\@.
With the $25\times5$ cell layout,\footnote{Here, we added 25\,\% buffer spaces (no qubits placed and hence no impact on qubit counts) for efficient routing of cells.} a single-step qubit rearrangement takes $\tau_r \approx 220~\mu$s, determined by the average time to route a cell with arbitrary initial and final locations in the grid.
We add a single step of syndrome extraction for $\tau_{SE} = 120~\mu$s after the routing round.
The length of a gate execution round is determined by the CNOT time, at $\tau_{CX} = 150~\mu$s followed by $\tau_{SE}$.
We include the routing of $\ket{Y}$ and $\ket{T}$ in the routing time $\tau_r$, so that the above concludes a single step of circuit layer execution, with a total time cost of $\tau_\text{cycle} = 610~\mu$s.

Single-trial factory runtime is $3$~ms for the chosen parameter set ($\tau_M = 100~\mu$s, $N_\mathrm{MB}=4$), where we achieve a minor improvement by reordering the auxiliary cells in Fig.~\ref{fig:factory-basic}(d) to eliminate two steps in the initial $CX$ rounds.
With $25$ factories available, we expect $5$ $T$ states to be generated during the execution of a single circuit layer; as such, we set total circuit depth to be $2\times 10^7$, since parallel execution of $T$ gates is naturally supported in our architecture.
With this assumption, the computation time for $100$ qubit, $10^8$ $T$-count circuit is $3.4$ hours.
For comparison, with a similar layout and fully pipelined syndrome qubit routing in the module assumed, as well as the same assumptions for the qubit performance, the equivalent time cost for the zoned approach is $16.5$ hours at a similar physical qubit count; see Appendix~\ref{app:zone} for the details of the assumed module layout and pipelined syndrome qubit routing.

Finally, we emphasize that the primary contribution of this work lies in the development of a theoretical framework that enables systematic resource estimation, as demonstrated here.
The present analysis is based on the above set of assumptions and abstracts away relatively minor factors, and more accurate resource estimation may require a deeper system-level analysis beyond the scope of this paper.
Nonetheless, our results point to potential directions for such investigations, which we outline in Appendix~\ref{sec:full-analysis}.
    
\section{Conclusion and outlook}\label{sec:concl}
In this paper, we have addressed a critical gap between the emerging transversal-gate implementations and the fundamental design requirements for scalable fault-tolerant quantum computers based on the surface code.
A crucial aspect of our work is to ensure the scalability of FTQC with reconfigurable qubits by allowing syndrome extraction even during the implementation of logical gates.
This feature makes it possible to scale up systems without violating the assumptions of the threshold theorem, which requires that syndromes be extracted frequently enough to keep error accumulation between extractions below a constant threshold.
We achieve this by moving away from the widely adopted zoned approach, where key physical operations rely on costly shuttling that becomes increasingly demanding as modules grow larger.
Instead, we propose using local gate operations and selective measurements---capabilities available in most reconfigurable qubit platforms---to enable syndrome extraction without the increasing time overhead associated with qubit shuttling.

Careful implementation design and instruction set architecture that respects these requirements resulted in a simple formulation of transversal-gate FTQC as the transversal surface-code game, played on a two-dimensional grid of cells with a finite set of rules governing mode transitions for the cells.
This establishes a distinct counterpart of the ``game of surface code'' framework~\cite{litinski2019game}, which was originally developed for superconducting qubit devices employing lattice surgery.
Our framework is designed to fully exploit the transversal gates, making it compatible with efficient techniques for transversal-gate FTQC, such as algorithmic fault tolerance. This highlights the fundamental difference in operational characteristics and design principles between these promising technologies and previous approaches.

We have also proposed a general method to design space-efficient resource state factories that are compatible with neutral atom systems.
These factories operate by employing transversal gates and pipelined shuttling of cells, rather than relying on multiple rounds of lattice-surgery Pauli product measurements.
This construction applies to various types of resource states, not only $\ket{T}$ and $\ket{Y}$ but also inter-module Bell pairs, which are essential for realizing multiprocessor fault-tolerant quantum computers at large scales~\cite{sunami2025scalable}.

Finally, the cost analysis within our game framework allowed us to perform resource estimation efficiently and systematically.
Notably, our evaluation based on near-term neutral-atom capabilities reveals that their space-time performance can be made comparable to a well-established baseline for superconducting qubits~\cite{litinski2019game}, owing to the use of time-efficient $SE$ together with techniques available for transversal-gate FTQC, thereby demonstrating the much-needed methodology to overcome the intrinsically slow physical operations of these systems by exploiting their unparalleled flexibility~\cite{saffman2025quantum,zhou2025resource}.
While a more thorough, full-system analysis is necessary for a fair comparison of vastly differing modalities (see also Appendix~\ref{sec:full-analysis}), this highlights the importance of comparing the quantum computer hardwares and fault-tolerant protocols fully optimized for their unique capabilities with great care, rather than relying solely on their physical operation timescales and qubit counts.

We expect that our framework serves as a useful abstraction and a systematic toolkit to interface with the rapid developments in experiments and theories. 
The systematic resource analysis is a crucial tool for evaluating the impact of particular experimental innovations,
while also serving as a framework to tailor novel theories for reconfigurable qubits.
We conclude this section by outlining several theoretical and experimental perspectives identified in this work, which we believe will motivate future works.
In Sec.~\ref{sec:qec-theory} we discuss pathways for incorporating other approaches for FTQC. 
This includes the incorporation of quantum error-correcting codes other than the rotated surface code, such as those with high encoding rates, while satisfying the fundamental requirement of the game.
In Sec.~\ref{sec:faster}, we point out the potential implications of the present analysis to the experimental efforts towards the development of neutral-atom FTQC.
Finally, in Sec.~\ref{sec:network}, we discuss the multiprocessor FTQC architecture and its incorporation into the game using high-speed quantum interconnect capabilities \cite{li2024high-rate, sunami2025scalable, sinclair2025fault-tolerant}.

    \subsection{Establishing theoretical frameworks for  scalable implementation of other fault-tolerant protocols}\label{sec:qec-theory}
    In this work, we have focused on the rotated surface code, where logical gate construction and error-correction performance are well established. 
    The two-dimensional local implementation for syndrome extraction is also essential as it enables fast execution using selective Rydberg gates, performed as needed in constant time.
    This capability is a crucial requirement for system scalability and for meeting the conditions set by the threshold theorem.
    However, fault-tolerant protocols based on the surface code incur relatively large spatial overhead; therefore, it is advantageous to explore codes with higher encoding rates~\cite{yamasaki2024time-efficient,yoshida2025concatenatecodessavequbits,tamiya2024polylogtimeconstantspaceoverheadfaulttolerantquantum,10.5555/2685179.2685184,8555154}.

    Indeed, in recent years, proposals for implementing high-rate QEC codes with shuttling-based interactions or long-range Rydberg interactions emerged, with a prospect for space overhead reduction~\cite{xu2024constant-overhead, viszlai2024matching, pecorari2025high-rate, poole2025architecture}.
    However, shuttling-based syndrome extraction, as well as selective long-range Rydberg interaction, incurs rapidly growing time costs to perform even a single round of $SE$ for larger code distances and system sizes, incurring significant time overhead. 
    Further, in some cases, efficient logical-gate operations are not yet well studied or inherently costly due to the nature of the high-rate codes that encode multiple logical qubits in a single code block.
    Most crucially, the increasing time cost of shuttling during syndrome extraction leads to a violation of the assumptions underlying the threshold theorem in these proposals, thereby limiting scalability and preventing the full exploitation of the potential of neutral atom platforms.

    For these fault-tolerant protocols to be compatible with neutral atom platforms, it is essential to develop a comprehensive theoretical framework that bridges the theory of FTQC with experimental realizations, as demonstrated in this work for the surface-code protocol. Such a framework must satisfy key requirements for scalable implementation, including the flexible execution of syndrome extraction necessary for the existence of a threshold, similar to our careful integration for the surface code, where cells are allowed to transition to the $SE$ mode at arbitrary times as needed.
    Furthermore, for a system-wide analysis, an efficient implementation of a universal gate set must be available for constructing the equivalent game structure with well-defined operation rules, which allows the space-time overhead analysis as performed in this work that correctly evaluates the overall benefit of these codes.
    The general design of the resource state factory in Sec.~\ref{sec:factory} allows us to straightforwardly incorporate new approaches for state distillation protocols, such as constant-overhead magic state distillation~\cite{wills2024constantoverheadmagicstatedistillation} and efficient distillation of remote entanglement~\cite{pattison2024fast, ataides2025constant-overhead}.
    Finally, it is interesting to establish a similar framework not only for the high-rate quantum low-density parity-check codes, but also for the high-rate concatenated codes~\cite{yamasaki2024time-efficient,yoshida2025concatenatecodessavequbits}.

    \subsection{Experimental perspectives}\label{sec:faster}
    
    In this work, we have proposed the use of a standalone cell with selective gates instead of a zoned approach, which is expected to resolve the fundamental scalability issue of neutral-atom systems.
    We believe this shift towards modular cell design and the associated systematic resource estimation capability supports more coordinated experimental developments to efficiently scale up the neutral-atom FTQC.

    In contrast to the zoned approach, where all physical operations relied on the globally shared zones with synchronization and coordination requirements, the standalone operation of the individual cell, made possible in our design, allows for an efficient technological development of neutral-atom systems.
    This is because the development can be separated into i) cell-level performance, such as the error-correction performance, logical gate implementation, and improvements in physical-gate operations, and ii) integrated grid-level performance, which includes coordinated operation of many cells, cell routing, quantum circuit compilation, factory operation, and system-wide control platform.
    In the zoned approach, long-distance atom shuttling was essential both in the individual logical block operation and system-wide operation, resulting in fundamental complications and trade-offs, which can be potentially avoided in our approach.
    The clear distinction of the operational components in the present architecture allows individual developments to have clear performance metrics while being able to be seamlessly combined to achieve a better overall performance.
    
    Previously, resource analysis of large-scale FTQC with concise abstraction of physical-level operation was not widely available for neutral-atom systems.
    With our approach, it is now possible to systematically connect the physical-level operational capability to the system-level performance metric, such as FTQC operation space-time overhead, presenting an opportunity to identify the bottleneck operation that may not have been apparent from the individual physical-level performance metrics routinely reported.
    Identifying the bottleneck factor of system-level performance is thus crucial for efficient and rapid experimental developments towards the realization of large-scale FTQC with neutral atoms.

    Finally, the resource analysis may also contribute to resolving the longstanding question of the neutral-atom approach regarding the optimal choice of atom species.
    While most atom species actively utilized for tweezer-array experiments are compatible with the present approach (see Sec.~\ref{sec:intro-atom}), their respective advantages and disadvantages affect the system-level performance in different ways.
    Systematic assessment of these characteristics and their contribution to the performance of large-scale quantum computing will be valuable for the experimental design and experimental effort allocation.

    \begin{figure*}[t]
        \centering
        \includegraphics[width=0.99\textwidth]{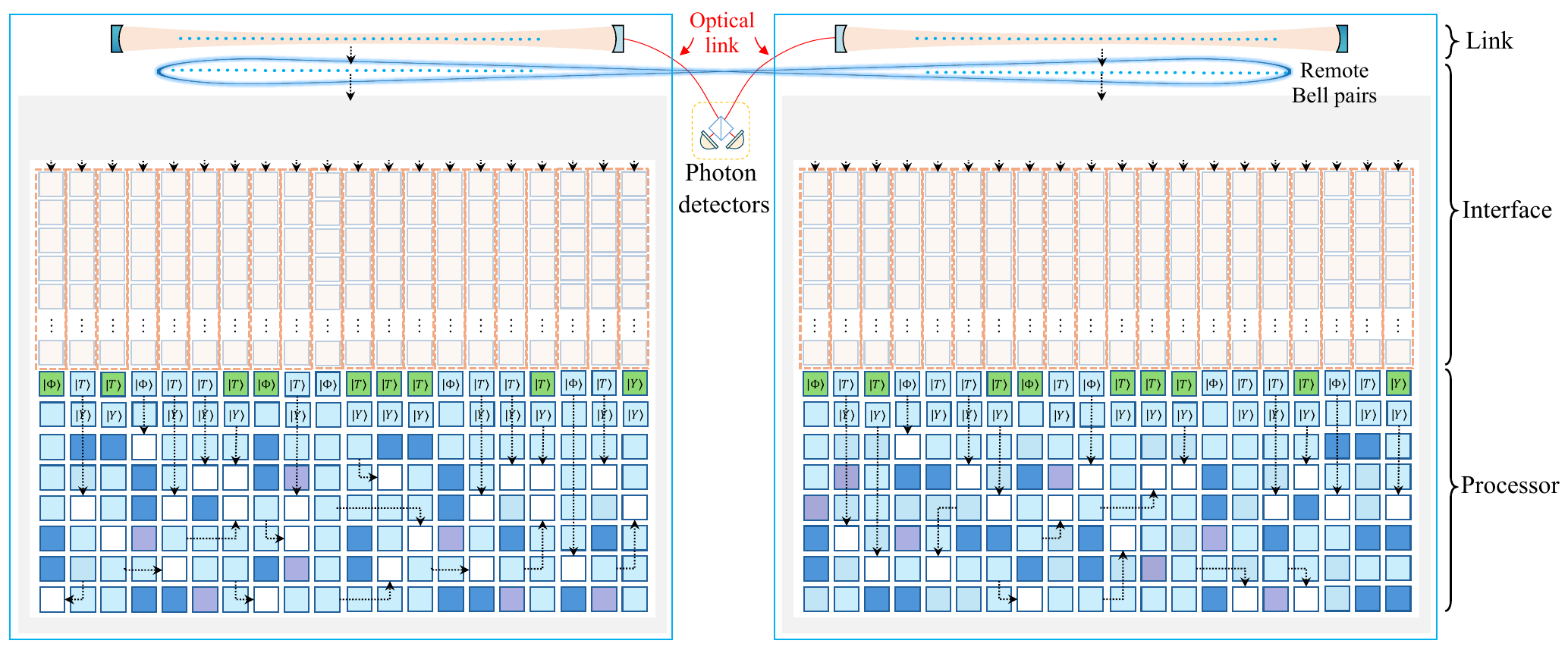}
        \caption{
        A vision for a multiprocessor fault-tolerant quantum computer, where multiple modules are equipped with interfaces to links that interconnect the modules.
        An optical interconnect device (optical cavity) is placed near the reloading zone, operating physical remote entanglement (Bell pair) generation.
        The entangled atoms are shuttled into the factory area, where logical entanglement distillation is performed to generate $\ket{\Phi}$ resource states.
        }
        \label{fig:processor_network}
    \end{figure*}
    
    \subsection{Toward Fully Scalable FTQC through Multimodule Neutral-Atom Architectures with Photonic Interconnects}\label{sec:network}
    
    Scaling up a single neutral-atom module beyond $10^4$--$10^5$-qubit order is challenging from both experimental and architectural points of view.
    Multiprocessor architecture is key to resolving this issue and achieving enhanced scalability~\cite{sunami2025scalable}, using entangled logical qubit pairs mediating interaction across modules.
    High-fidelity resource state generation protocols such as logical entanglement distillation~\cite{sunami2025scalable, pattison2024fast, ataides2025constant-overhead} and lattice surgery-based entangled logical qubit pair generation~\cite{ramette2024fault-tolerant} are promising approaches.
    High-rate, high-fidelity physical-level optical interconnect protocols are actively developed to support efficient logical level implementation, which requires a large number of entangled qubit pairs;  entanglement generation rate of over $100$ kHz is expected to be achieved for neutral-atom qubits by the use of near-term cavity quantum electrodynamics hardware~\cite{sunami2025scalable, li2024high-rate, huie2021multiplexed, kikura2025taming, kikura2025passive}.
    Their incorporation into the present architecture will enable system-level evaluation to identify ideal interconnect performance metrics.
    For example, physical-level optical link devices can be incorporated as part of the qubit buffer, supplying entangled qubit pairs to pipelined entanglement distillation factories.
    In Fig.~\ref{fig:processor_network}, we illustrate a vision for a fully scalable multiprocessor approach to building a large-scale fault-tolerant quantum computer; toward this goal, our framework provides both a theoretical foundation and a practical toolkit for developing and analyzing such architectures.

\section*{Declaration of Competing Interest}
S. Sunami and H. Yamasaki are employees, and A. Goban is a co-founder and a shareholder of Nanofiber Quantum Technologies, Inc.

\begin{acknowledgements}
We thank Y. Suzuki and Y. Ueno for discussions, and I. Byun, R. Inoue, and V. Vuleti\'c for providing feedback on the manuscript. 
We thank the NanoQT team for providing the environment in which this work was possible.
\end{acknowledgements}

\appendix

\setcounter{table}{0}
\setcounter{figure}{0}
\renewcommand{\thetable}{A\arabic{table}}
\renewcommand{\thefigure}{A\arabic{figure}}
\renewcommand{\theHtable}{Supplement.\thetable}
\renewcommand{\theHfigure}{Supplement.\thefigure}

\newpage
\section*{Appendices}

Appendices are organized as follows.
In Appendix \ref{app:local}, we discuss the state-of-the-art performance of local gates and measurements with neutral atoms, as well as their near-term prospects.
In Appendix \ref{app:operation}, we describe the basis of shuttling time cost evaluation, including the array rotation, and also provide an estimated time cost for the fold-transversal $S$ gate execution time based on the protocol of Ref.~\cite{chen2024transversal}.
We discuss pathways for incorporating the active qubit-loss correction subroutines into the proposed approach in Appendix \ref{sec:loss-correction}.
In Appendix \ref{app:zone}, we describe a baseline zone layout for the alternative resource estimation for the zone-based approach reported in Sec.~\ref{sec:example}.
Finally, in Appendix \ref{sec:full-analysis}, we discuss several approaches to performing thorough resource estimations with detailed cost analysis.

\section{Local operations of neutral-atom array}\label{app:local}
Here we summarize the current and near-term experimental perspectives of selective gate and measurement operations with neutral-atom arrays, which are the basis of the proposed architecture. 
We first discuss the selective gate implementation and its viability in Sec.~\ref{app:selective-cz}, and selective, mid-circuit measurement in Sec.~\ref{app:selective-meas}.

\subsection{Selective Rydberg CZ gates}\label{app:selective-cz}

State-of-the-art Rydberg $CZ$ gates now exceed $99\,\%$ fidelity across multiple elements—$99.71\,\%$ in Sr~\cite{scholl2023erasure}, $99.4\,\%$ in Yb~\cite{peper2025spectroscopy, muniz2024high-fidelity}, $99.5\,\%$ in Rb~\cite{bluvstein2024logical}, and $99.35\,\%$ for individually addressed Cs qubits~\cite{radnaev2025universal}—with gate times of $\approx100\,$ns.
Crosstalk from long-range van-der-Waals tails remains below $10^{-4}$ when the integrated stray shift on each qubit stays under $1\,\%$ of the inverse pulse duration~\cite{poole2025architecture}, a condition satisfied by spacing neighboring $CZ$ pairs at least two lattice sites apart and keeping the Rydberg beam waist below half the lattice period~\cite{radnaev2025universal}.
Under these parameters, the gates in $SE$ can be completed in $\tau_\mathrm{SE}^{(\mathrm{gate})}\!\approx\!20~\mu\mathrm{s}$ (see also Sec.~\ref{sec:se}); the dominant time overhead is the $\approx1~\mu\mathrm{s}$ AOD switching time required to reconfigure the addressing beams, not the Rydberg interaction itself~\cite{Graham2023selectiveaddressing, radnaev2025universal}.
Photonic-integrated beam-steering arrays with sub-10 ns reconfiguration could reduce this overhead by an order of magnitude~\cite{Menssen2023picbeamshaping},
with their spatially multiplexed implementation providing an avenue for relaxed laser-power requirements and laser-delivery complexities. 

\subsection{Local measurements}\label{app:selective-meas}
We use local measurements implemented without a dedicated detection zone by exploiting atomic-species or internal-state selectivity.  
In a dual-species register, for example, syndrome atoms fluoresce at wavelengths far off-resonant for the data species, so scattered photons leave the data qubits effectively dark.  
Experiments with Rb–Cs arrays~\cite{singh2023mid-circuit,fang2025interleaveddualspecies}, dual-isotope arrays~\cite{Sheng2022Rbdualspecies,nakamura2024dualisotope}, and the optical–metastable–ground manifold of ${}^{171}\mathrm{Yb}$~\cite{lis2023midcircuit} demonstrate mid-circuit readout with neighboring-qubit disturbance below $10^{-4}$.

The duration of these measurements is now the principal overhead in the $SE$ cycle.  
Recent demonstrations of non-destructive detection in tweezer arrays have achieved \(\approx160~\mu\mathrm{s}\)~\cite{shea2020submillisecond}, as well as $\approx 250 \mu\mathrm{s}$ for higher fidelity of 99.9\%~\cite{chow2023high-fidelity}, with prospects for rapid improvements by advancement of optical systems and detecting devices.  
High-intensity lattice imaging has recently reached \(2.4~\mu\mathrm{s}\) with \(99.4\,\%\) fidelity~\cite{su2025fastimaging}, pointing to a clear path toward sub-\(100~\mu\mathrm{s}\), high-fidelity mid-circuit measurements for tweezer-trapped atoms (see also Sec. \ref{sec:faster}).

\section{Time cost evaluations}\label{app:operation}
In this section, we describe the detailed steps for evaluating the time costs, including the fold-transversal gates for the rotated surface code in Ref.~\cite{chen2024transversal}.
We first present the analytical model for evaluating the shuttling time cost for a given distance in Sec.~\ref{app:STA}, based on Ref.~\cite{hwang2024fast}, which is the basis for the cost analysis in Sec.~\ref{sec:operating}.
In Secs.~\ref{app:rotation-time} and \ref{app:fold-s-time}, we discuss the estimated gate times based on the shuttling model for $H$ and $S$ gates, respectively.

\begin{figure}[t]
    \centering
    \includegraphics[width=0.75\linewidth]{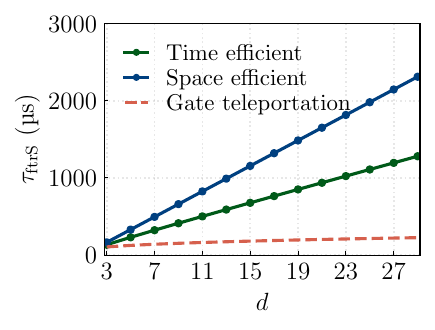}
    \caption{Time cost of fold-transversal implementation of an $S$ gate to the mid-cycle states of rotated surface code with STA-based atom shuttling. We assume the lattice constant of the data-qubit array to be $5 \mu$m.
    For comparison, given $\ket{Y}$, the time cost of gate teleportation of an $S$ gate is dominated by that of $CX$, i.e., $\tau_{CX}$, which is evaluated as in Fig.~\ref{fig:times} and incorporated in the figure as a dashed line for reference.
    Even within the regime shown in this plot, gate teleportation offers a time advantage of an order of magnitude compared to the fold-transversal implementation.
    }
    \label{fig:ftrS}
\end{figure}

\begin{figure*}[t]
    \centering
    \includegraphics[width=0.99\textwidth]{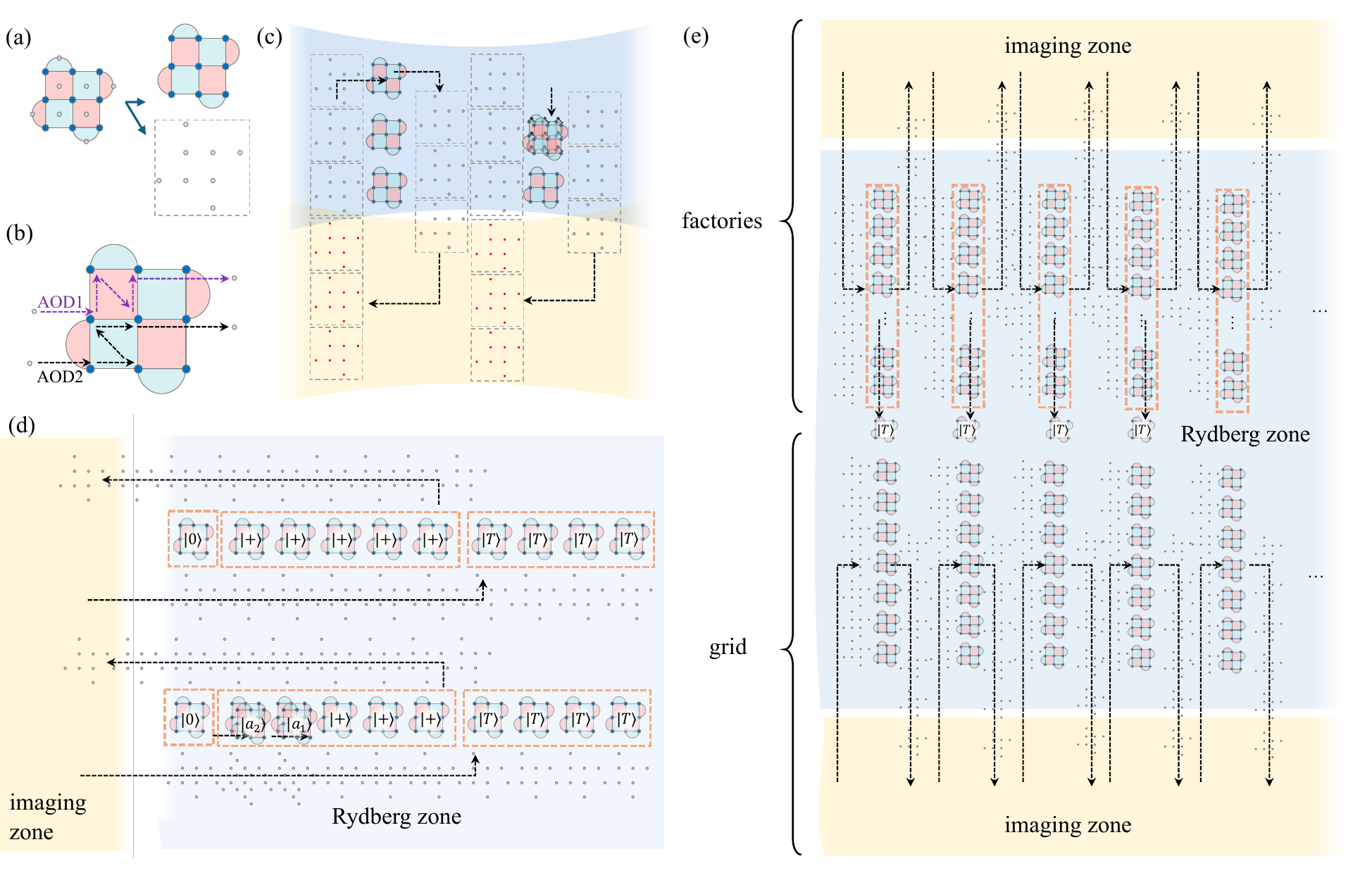}
    \caption{A reference design for zoned approach, based on the processor design of this paper. 
    (a) In zoned implementation, data qubit and syndrome qubits are separately operated, which results in variable time cost for operating the syndrome extraction.
    (b) Pipelined shuttling-based error correction requires the syndrome qubits to be brought into the logical qubit array, interacted with the data qubits before being shuttled out of the array.
    (c) An example layout for a fully pipelined syndrome qubit operation. Depending on the measurement time and shuttling times, the required number of additional syndrome qubits varies.
    (d) Factory design of Fig.~\ref{fig:factory} adapted to zoned approach with pipelined syndrome qubit operation.
    (e) Overall module layout with factory and grid with dedicated imaging zones. 
    }
    \label{fig:zone}
\end{figure*}

\subsection{Atom shuttling for cell routing}\label{app:STA}
Major costs for atom shuttling arise from the need to move the trap while maintaining the motional states of the atoms in the trap, with constant-jerk (CJ) trajectory being the widespread choice.
A fast alternative using shortcuts to adiabaticity (STA) is recently demonstrated in several experiments \cite{hwang2024fast, Cicali2024statransportoptimization, chinnarasu2025variational}.
Analytical expressions for the motional excitation for shuttling distance $l$, atomic mass $m$, shuttling time $t_f$, and trap angular frequency $\omega_0$ are given as \cite{hwang2024fast}:

\begin{eqnarray}
\centering
\Delta n \approx
\begin{cases}
\displaystyle \frac{m l^2}{2 \hbar \omega_0} \times t_f^{-2} & \text{(CV)}, \\
\displaystyle \frac{36 m l^2}{\hbar \omega_0^3} \times t_f^{-4} & \text{(CJ)}, \\
\displaystyle \frac{3600 m l^2}{\hbar \omega_0^5} \times t_f^{-6} & \text{(STA)},
\end{cases}
\end{eqnarray}
where CV denotes a constant-velocity trajectory. 
Depending on the implementation, we may need to add adiabatic transfer between stationary and dynamic optical tweezers before and after the movement, which takes additional $\approx 40$ $\mu$s with STA trajectory \cite{Cicali2024statransportoptimization}, while it is also possible for the atoms to be trapped in movable tweezers throughout relevant operations; as such, we neglect this cost for the estimations in the main text.
Further, for scalable time-cost analysis, we restricted ourselves to the case where routing-based CNOT gate must be performed only between adjacent logical qubits, by dividing long-distance routing and CNOT gate operations into separate modes. In practice, cells at a relatively large distance may perform transversal CNOT gates in a single step, as long as the error accumulation and the time cost for routing remain negligible or within the allowed budget. We leave such optimization to future work.
For the time cost evaluation of Fig.~\ref{fig:times}, we use $\Delta n = 1$ as the motional excitation threshold for the atom shuttling.
Then, we obtain the time cost $t_f$ for a given distance $l$ based on the expression above, using typical parameters for neutral atom systems, specifically, $\omega_0/2\pi = 100$ kHz and $m$ of ${}^{171}$Yb\@.\footnote{Lighter atoms yield slightly shorter transport times; while many species of atoms actively utilized for tweezer array experiments, also including lighter species such as ${}^{87}$Rb, are compatible with the proposed approach, the choice of atom species in practice should be based on careful consideration of overall performance, not just transport times, as discussed in Sec.~\ref{sec:faster}.}

\subsection{Cell rotation for $H$ gates}\label{app:rotation-time}
Apart from our proposal on direct cell rotation in implementing logical $H$ gates on the surface code, a conventional method for cell rotation is based on AOD (see also Fig.~\ref{fig:times}).
For the estimation of  AOD-based cell rotation, we follow the horizontal-diagonal reflection method of Ref.~\cite{chen2024transversal}, with the divide-and-conquer strategy of Ref.~\cite{xu2024constant-overhead}.
This involves $\lceil\log_2(d)\rceil+1$ steps of atom shuttling for the reflection, with a maximum shuttling distance of $d$ lattice sites.
For a time-efficient protocol, 50\,\% larger area is used for the temporary storage of atoms, while a space-efficient protocol requires a factor of 2 larger time cost to maintain the array inside the cell area.
Shuttling of tweezer-trapped atoms along a circular path was also demonstrated in Ref.~\cite{hwang2024fast}, which confirmed the analytical expression provided in Ref.~\cite{hwang2024fast}. Again, we use $\Delta n = 1$ as the motional excitation threshold for the atom shuttling.
In  Fig.~\ref{fig:times}(b), we use this heating threshold to plot the time cost for the outermost qubits in the array to achieve the circular path, with less heating to the other qubits in the cell being rotated together.

\subsection{Fold-transversal implementation of $S$ gates}\label{app:fold-s-time}
Instead of teleporting $S$ gates used in our approach, fold-transversal implementation of $S$ gates for the rotated surface code is also common, achieved by operating on mid-cycle states during the SE~\cite{chen2024transversal}. 
While this implementation has good performance in terms of logical gate errors at the scales currently available in experiments~\cite{chen2024transversal}, we propose to use gate teleportation with transversal gates due to their time efficiency and compatibility with the requirements of the game.

In Fig.~\ref{fig:ftrS}, we show the estimated time cost to perform the fold-transversal implementation of $S$ gates using diagonally aligned two-dimensional AODs with STA trajectory.
The time-efficient protocol utilizes additional space to allow parallel shuttling of the atom array, while the space-efficient protocol is designed to maintain qubits inside the cell area.
For comparison, as discussed in Sec.~\ref{sec:tel-gate}, the gate teleportation of $S$ is implementable with a time cost dominated by that of $CX$, i.e., $\tau_{CX}$, where the resource state $\ket{Y}$ can be efficiently prepared using the optimized factories in Sec.~\ref{sec:y-factory}.
The figure shows that gate teleportation of $S$ offers an order-of-magnitude time advantage over the fold-transversal implementation, even at small scales.

\section{Incorporating qubit-loss correction subroutines}\label{sec:loss-correction}
    Qubit loss is a critical factor for operating large-scale quantum computation with neutral atoms, induced in a wide variety of operations, including idling, gates, and measurements. 
    Several loss-correction protocols are available for tackling this issue, such as the measurement-based approach \cite{yu2025processing}, 
    Knill's teleportation-based error correction~\cite{knill2005quantum}, loss-detection unit \cite{perrin2025quantum, chow2024circuit-based}, and swap-incorporated syndrome extraction \cite{suchara2015leakage, baranes2025leveraging}, with comparative analysis for neutral-atom systems recently performed in Ref.~\cite{baranes2025leveraging}.

    Incorporating the above approaches to the transversal surface-code game is possible by either absorbing them into existing modes, such as the $SE$ mode, while carefully maintaining the constant-time, flexible syndrome extraction capability, or by defining additional modes for loss correction with associated mode transition rules.
    An outstanding challenge is the requirement of a scalable, constant-time protocol for refilling lost tweezer sites with atoms. 
    Since the atom reservoir is typically available only at fixed locations of the grid, such as the boundary of the grid (see, e.g., Fig.~\ref{fig:processor}), direct refilling from the reservoir results in variable and system-size dependent shuttling time that violates the requirement for the threshold theorem and the rule of the game.
    
    As a potentially viable approach for achieving constant-time atom supply, we propose a relayed refill scheme: an atom lost from a site in a cell is supplied from the neighboring cell, which in turn obtains an atom from its own neighbor, relaying the atoms until a cell adjacent to the reservoir finally refills directly from the reservoir, completing the relay chain.
    We leave the detailed formulation of game rules incorporating this scheme and its performance evaluation for future work.

\section{Zoned Approach for FTQC}\label{app:zone}
In this section, we outline the baseline zoned approach used as a point of comparison with our approach.
A challenge in this comparison is that the performance of zoned architecture is expected to vary significantly based on the zone arrangement.
For a sufficiently fair comparison, we strived to design the module layout as close to the one in Fig.~\ref{fig:processor} as possible, with the same resource state factory layout and fully pipelined syndrome qubit routing. 
We show the assumed zone and logical-qubit layout in Fig.~\ref{fig:zone} (e): we placed the syndrome-qubit pipeline in between columns of the logical qubit array to achieve minimal time cost for the $SE$ as plotted in Fig.~\ref{fig:times}(d). 
In contrast to the estimation based on our game-based approach, it is possible to omit the qubit measurement times from the SE time cost for the zoned approach since they are performed in parallel in the measurement zone.
The use of pipelined syndrome qubits incurs qubit-count overhead to prepare a sufficient number of ancilla qubits to be shuttled across the zones continuously for repeated error correction while avoiding wait time in the grid, with at least a factor of three increase in syndrome qubit count for the assumed parameter regime.
This means that, for a single logical qubit with $n=d^2$ data qubits, at least $3(n-1)$ syndrome qubits are required, increasing the space overhead by a factor of two.
To keep the total physical qubit count similar to Table \ref{table}, we thus need to reduce the number of  $\ket{Y}$ factories to 25, and $\ket{T}$ factories to $8$, reducing the $\ket{T}$ throughput to $1.5$ per $\tau_\text{cycle}^\text{(zone)} = 900$~$\mu$s. 
With the required depth to execute $10^8$ $T$ gates being $6.6\times 10^7$, the total runtime is 16.5 hours.

\section{More detailed overhead analysis}\label{sec:full-analysis}

    In this section, we discuss several pathways for a more comprehensive resource estimation, incorporating the implementation-specific overheads.
    
    Qubit reuse and pipelined routing: in our estimation in Sec.~\ref{sec:example}, we omitted the qubit-count overhead for a pipelined routing of moving cells in the factories; after being measured in the measurement area, they need to be routed back to the feed area for efficient reuse, e.g.~through the gray shaded region in Fig.~\ref{fig:processor}.
    The same requirement applies to the $\ket{Y}$ states used in the factory, which need to be buffered after the interaction and measured, before being routed back to $\ket{Y}$ factories for qubit reuse.
    If implemented efficiently, the fundamental limiting factor is the shuttling from the measurement area to the feed of the factory, at $< 200 \mu$s for the parameters in Table~\ref{table}; with fully pipelined operation, for example, using routing space between factories, this overhead can be negligible as this timescale is compatible with the single step of pipelined operation in the factory.
    Similarly, qubit reuse after gate teleportation in the grid must be considered: they need to be routed from the grid to the routing buffer (gray shaded region of Fig.~\ref{fig:processor}), to the feed area of the factories. 
    Our architecture utilizes a one-way supply of new qubits, from top to bottom in Fig.~\ref{fig:processor}, with the routing buffer (gray shaded area) used for moving the qubits back to the feed of the factories; thus, it is expected to be pipelined efficiently albeit with potentially some spatial overhead.
    An additional overhead may come from the measurement and sorting of the atom array before being fed into the factory to ensure a defect-free array.

    Direct atom-array rotation: the direct-rotation-based logical Hadamard gate implementation we assumed in Sec.~\ref{sec:example} rests on the ability to perform atom-array rotation with STA trajectory. 
    As discussed in the main text, there are promising options to perform this task; however, actual implementation may incur additional space and time overhead.
    A particular challenge is the effect of rotating the tweezer array on nearby cells, which must be sufficiently suppressed by spatial separation or a dedicated region for rotation.
    Fundamentally, our estimation targeted to a fixed $T$-count is primarily bottlenecked by the $\ket{T}$ factory throughput, and our factory does not use the $H$ gate (see Fig.~\ref{fig:factory}); thus, the effect of increased cost of this gate is expected to be mild, while the throughput of the catalytic $\ket{Y}$ factories, as well as space-time overhead of operation in the grid, will be directly affected, resulting in a possible difference in overall performance. 
    
    Cell sorting rounds: In our estimation, we assume only a single routing round between each circuit execution layer. 
    While a combination of efficient routing algorithms and tailored circuit compilation may allow such an operation, especially for highly structured circuits found in many quantum algorithms, it is challenging to guarantee for general quantum circuits; as such, in general, multiple rounds of cell routing may be required.
    Again, since our estimates are bottlenecked by the $\ket{T}$ factory throughput, operating independently of the grid, and since our factory output is placed along an edge of the grid for efficient routing to any target cell, the effect of increased routing rounds on the overall time cost is also expected to be limited.
    Nevertheless, optimized cell shuffling and circuit compilation strategies are essential for the efficient implementation of large-scale FTQC with neutral atoms employing transversal gates.

\end{document}